\documentclass[twocolumn,twocolappendix]{aastex631}
\usepackage{rotating}
\usepackage [english]{babel}
\usepackage [autostyle, english = american]{csquotes}
\MakeOuterQuote{"}
\pdfoutput=1 
\usepackage{graphicx}
\usepackage[inline]{enumitem}
\usepackage{verbatim}
\usepackage{color}
\usepackage{longtable}

\hypersetup{pdfauthor={Ewan O'Sullivan},
            pdftitle={HCG 57: Evidence for shock-heated intergalactic gas from X-rays and optical emission line spectroscopy},
            pdfkeywords={galaxy groups; Hickson compact group; galaxy interactions; galaxy collisions},
            bookmarksnumbered=true}

\newcommand{\Hi}{\ion{H}{1}}

\newcommand{\cii}{[\ion{C}{2}]}
\newcommand{\oi}{[\ion{O}{1}]}
\newcommand{\nii}{[\ion{N}{2}]}
\newcommand{\kms}{km~s$^{-1}$}
\newcommand{\arcm}{\hbox{$^\prime$}}
\newcommand{\degree}{\hbox{$^\circ$}}
\newcommand{\chandra}{\emph{Chandra}}
\newcommand{\xmm}{\emph{XMM-Newton}}
\newcommand{\xmms}{\emph{XMM}}
\newcommand{\arcs}{\mbox{\arcm\arcm}}
\newcommand{\Zsol}{\ensuremath{\mathrm{~Z_{\odot}}}}

\newcommand{\Msol}{\ensuremath{\mathrm{~M_{\odot}}}}
\newcommand{\Msolpyr}{\ensuremath{\mathrm{~M_{\odot}~yr^{-1}}}}
\newcommand{\Dtf}{\ensuremath{D_{\mathrm{25}}}}
\newcommand{\s}{\ensuremath{\mbox{~s}}}
\newcommand{\ps}{\ensuremath{\s^{-1}}}
\newcommand{\cm}{\ensuremath{\mbox{~cm}}}
\newcommand{\pcmsq}{\ensuremath{\cm^{-2}}}

\newcommand{\kev}{\ensuremath{\mbox{~keV}}}
\newcommand{\kevcmsq}{\ensuremath{\kev\cm^{2}}}
\newcommand{\km}{\ensuremath{\mbox{~km}}}

\newcommand{\erg}{\ensuremath{\mbox{~erg}}}
\newcommand{\ergps}{\ensuremath{\erg \ps}}

\newcommand{\kmps}{\ensuremath{\km \ps}}

\received{---}
\revised{---}
\accepted{---}
\submitjournal{ApJ}

\shorttitle{Shocked gas in HCG~57}
\shortauthors{O'Sullivan et al.}

\begin{document}

\title{HCG~57: Evidence for shock-heated intergalactic gas from X-rays and optical emission line spectroscopy}

\author[0000-0002-5671-6900]{Ewan O'Sullivan}
\affiliation{Center for Astrophysics $|$ Harvard \& Smithsonian, 60 Garden Street, Cambridge, MA 02138, USA}

\author[0000-0002-7607-8766]{P. N. Appleton}
\affiliation{Caltech/IPAC, MC 6-313, 1200 E. California Blvd., Pasadena, CA 91125, USA}

\author[0000-0002-7593-8584]{B.~A.~Joshi}
\affiliation{William H. Miller III Department of Physics and Astronomy, Johns Hopkins University, Baltimore, MD 21218, USA}

\author[0000-0002-3249-8224]{Lauranne Lanz}
\affiliation{Department of Physics, The College of New Jersey, 2000 Pennington Road, Ewing, NJ 08628, USA}

\author[0000-0002-4261-2326]{Katherine Alatalo}
\affiliation{Space Telescope Science Institute, 3700 San Martin Dr., Baltimore, MD 21218, USA}
\affiliation{William H. Miller III Department of Physics and Astronomy, Johns Hopkins University, Baltimore, MD 21218, USA}

\author[0009-0007-0318-2814]{Jan M. Vrtilek}
\affiliation{Center for Astrophysics $|$ Harvard \& Smithsonian, 60 Garden Street, Cambridge, MA 02138, USA}

\author[0000-0001-8952-676X]{Andreas Zezas}
\affiliation{Physics Department, University of Crete, Heraklion, Greece}
\affiliation{Center for Astrophysics $|$ Harvard \& Smithsonian, 60 Garden Street, Cambridge, MA 02138, USA}
\affiliation{Institute of Astrophysics, Foundation for Research and Technology-Hellas, N. Plastira 100, Vassilika Vouton, 71110 Heraklion, Greece}

\author[0009-0003-9413-6901]{Laurence P. David}
\affiliation{Center for Astrophysics $|$ Harvard \& Smithsonian, 60 Garden Street, Cambridge, MA 02138, USA}

\begin{abstract}
We present \chandra\ and \xmm\ X-ray observations of the compact group HCG~57, and optical integral field spectroscopy of the interacting galaxy pair HCG~57A/D. These two spiral galaxies recently suffered an off-axis collision with HCG~57D passing through the disk of A. We find evidence of a gas bridge linking the galaxies, containing  $\sim$10$^8$\Msol\ of hot,  $\sim$1~keV thermal plasma and warm ionized gas radiating in H$\alpha$, H$\beta$, [\ion{O}{3}] and \nii\ lines. The optical emission lines in the central regions of HCG~57D show excitation properties consistent with \ion{H}{2}-regions, while the outer rim of HCG~57D, parts of the bridge and the outer regions of HCG~57A show evidence of shocked gas consistent with shock velocities of 200-300~km~s$^{-1}$. In contrast, the X-ray emitting gas requires a collision velocity of ~650-750~km~s$^{-1}$ to explain the observed temperatures. These different shock velocities can be reconciled by considering the contributions of rotation to  collision velocity in different parts of the disks, and the clumpy nature of the pre-shock medium in the galaxies, which likely lead to different shock velocities in different components of the turbulent post-shocked gas. We examine the diffuse X-ray emission in the group members and their associated point sources, identifying X-ray AGN in HCG~57A, B, and D. We also confirm the previously reported $\sim$1~keV intra-group medium and find it to be relaxed with a low central entropy (18.0$\pm$1.7\kevcmsq\ within 20~kpc) but a long cooling time (5.9$\pm$0.8~Gyr).

\end{abstract}


\section{Introduction}

Galaxy groups provide an ideal laboratory in which to study a fundamental process in the evolution of galaxies and galaxy systems: the transformation of galaxies from cold-gas-rich, star-forming spirals to cold-gas-poor, ``red and dead'' ellipticals and S0s \citep[e.g.,][]{Bitsakisetal10,Bitsakisetal11,Bitsakisetal16}. The low velocity dispersions and small galaxy separations in groups make harassment and galaxy mergers common, and they are thought to play a central role in galaxy evolution. \Hi\ studies of compact groups (CGs) suggest an evolutionary sequence in which galaxy interactions strip cold gas from spiral galaxies to form intergalactic clouds and filaments, or even a cold intra-group medium (IGrM) \citep{Verdes-Montenegroetal01,Johnsonetal07,Borthakuretal10,Konstantopoulosetal10}. X-ray observations of groups have highlighted a number of processes which may play an important role: a) ram-pressure and viscous stripping of cold gas as spirals move through a hot IGrM \citep{Rasmussenetal06a,Rasmussenetal08}; b) tidally-induced starbursts which can eject significant masses of hot gas into the IGrM, e.g., in HCG~16 \citep{OSullivanetal14b,OSullivanetal14c}; and c) collision shocks, which can directly heat \Hi\ into the X-ray phase, e.g., in Stephan's Quintet \citep[HCG~92,][]{Trinchierietal03,Trinchierietal05,OSullivanetal09} and the Taffy galaxies \citep[UGC~12914/5,][]{Appletonetal15}.

Infrared observations provide a complementary window on the processes driving galaxy transformation. A \textit{Spitzer} spectroscopic survey of galaxies in CGs with intermediate \Hi\ deficiencies \citep{Cluveretal13} shows that many have suppressed star formation rates (SFRs), coupled with strongly enhanced H$_2$ emission, well above that expected for heating by young stars or X-rays. The extreme H$_2$ luminosities of these molecular hydrogen emission galaxies (MOHEGs) are powered by shocks and turbulence caused by the mechanical energy injected by galaxy collisions, or in some cases by radio jets \citep{Ogleetal10}. Shocks also drive enhanced emission in other lines, e.g., \cii\ in the far-IR \citep{Appletonetal13,Alataloetal14}. In some systems, suppressed star formation has been observed \citep{Alataloetal15,Lanzetal16}. Turbulent heating of the molecular gas has been suggested as a reason this, able to render the gas far more inefficient at turning into stars than would be expected \citep{Kennicutt98,Salimetal20}.

Considered together, these results suggest that low mass CGs are at a critical evolutionary stage, with collisions and tidal interactions driving both galaxy transformation and the build-up of the hot IGrM. Very few low mass CGs have high quality X-ray and infrared data available, but the power of a multi-wavelength approach is demonstrated by two famous systems. In Stephan's Quintet, an 850~\kms\ collision between an infalling galaxy and an intergalactic cold gas filament has shock heated $>$5$\times$10$^8$\Msol\ of \Hi\ into the X-ray regime, producing a bright, clumpy ridge of $\sim$0.6~keV X-ray emission linking the remaining \Hi\ clouds. However, detections in H$\alpha$ \citep{Sulenticetal01}, H$_2$ \citep{Appletonetal06}, CO \citep{Guillardetal12}, \cii\ and even H$_2$O \citep{Appletonetal13} show that both warm gas and cold dense clouds persist in the shock zone.  Modelling of the H$_2$ emission \citep{Appletonetal17} shows that a turbulent cascade transfers the energy of the collisional shock down into the coolest gas, where it is efficiently radiated away \citep{Guillardetal09,Guillardetal12}. Molecular hydrogen emission is in fact the dominant form of radiative cooling in the shock region; L$_{H_2}$ exceeds L$_X$ by at least a factor of 3 \citep{Cluveretal10}. Similar conditions are seen in the Taffy galaxies, a pair of spirals which have undergone a face-on collision, passing through each other's disks, drawing out a 20~kpc bridge of gas and tangled magnetic fields between them. This bridge contains a complex mixture of cool (CO, \Hi, H$_2$) and hot gas (X-ray, radio continuum emission) and shows multiple signatures of shock heating. The H$_2$ luminosity of the bridge, L$_{H_2}$$\sim$42$\times$L$_X$ \citep{Appletonetal15}, is exceptionally high, even compared with star-forming or AGN jet-interaction systems \citep{Lanzetal15}.

\begin{figure}
\centering
\includegraphics[width=\columnwidth,bb=55 55 555 735]{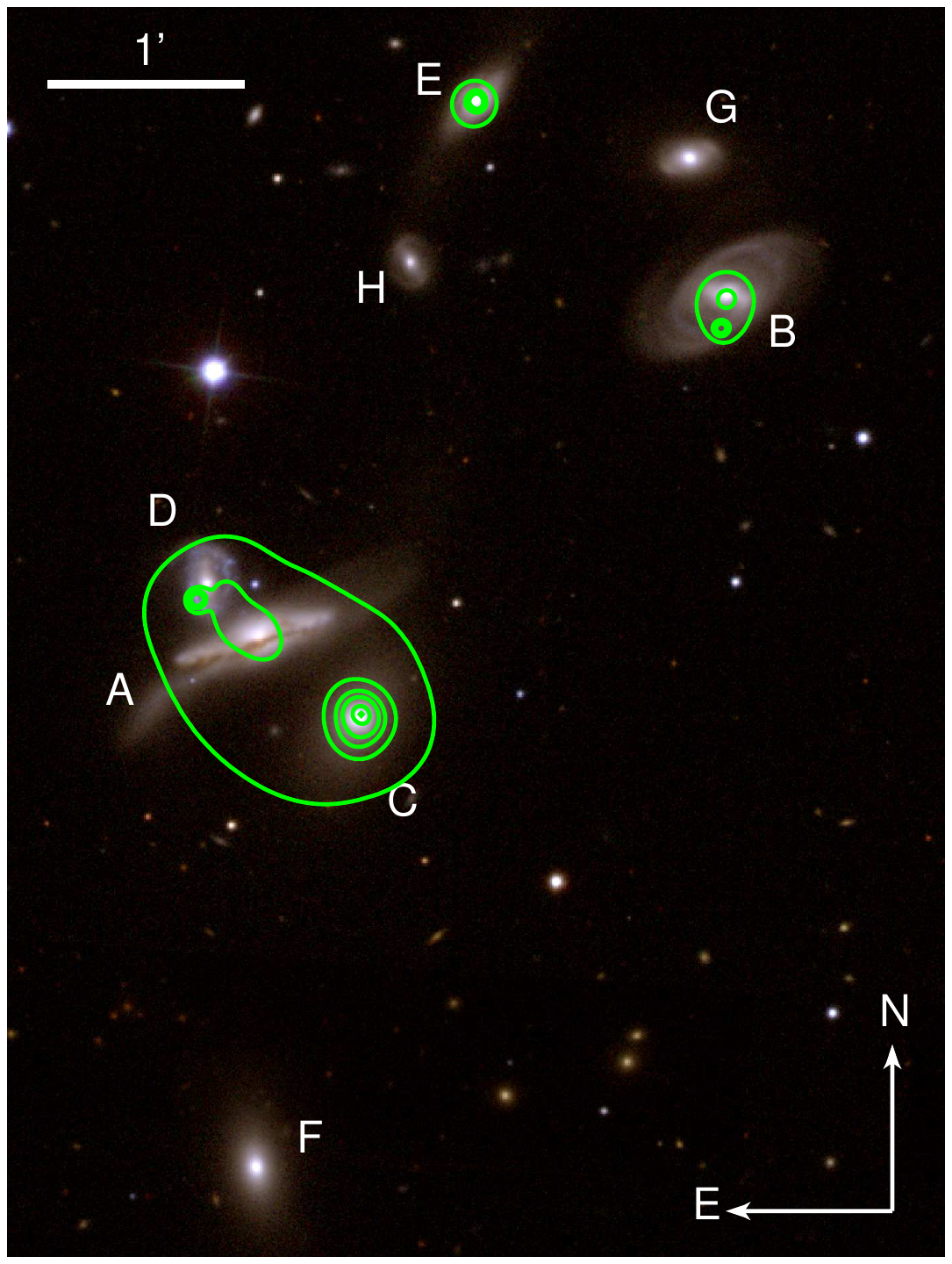}
\caption{\label{fig:SDSS}SDSS false-color image, using the $g$, $r$ and $i$ bands. The principal galaxies of the group are labelled, and adaptively smoothed \chandra\ X-ray contours are overlaid. The brightest X-ray emission is found in and around HCG~57A, C and D in the group core, with bright compact sources visible in HCG~57B and E.}
\end{figure}

In this paper we describe deep X-ray and optical integral field unit (IFU) observations of a compact group which hosts a similar interaction, HCG~57 (shown in Figure~\ref{fig:SDSS}). The group \citep[also known as the Copeland Septet or ARP~320,][]{DeVaucouleursetal76,Arp66} consists of 8 galaxies \citep{Hicksonetal89} with velocities 8700-9600\kmps, dominated by a triplet consisting of two spiral galaxies (HCG~57A \& D) and one elliptical (HCG~57C). \citet{Alataloetal14} argue that the peculiar kinematics of galaxies A \& D indicate that an off-center collision has occurred, in which D has passed through the disk of A, creating ring-like morphologies in both. Powerful warm H$_2$ emission was identified from the edge-on spiral HCG~57A which can only be explained by heating of the molecular gas by shocks \citep{Cluveretal13}. Follow-up observations by \textit{Herschel} (\cii~157$\mu$m, \oi~63$\mu$m), IRAM 30m and CARMA CO (1-0) mapping showed very broad and disturbed emission lines in the diffuse atomic and molecular gas, suggesting turbulent motions \citep{Alataloetal14,Lisenfeldetal14} with shocks possibly enhancing the \cii\ emission, as in Stephan's Quintet. Interestingly, a significant difference in star formation rates is observed between the smaller HCG~57D \citep[the most strongly star forming group member, SFR=1.77\Msolpyr,][]{Bitsakisetal11} and the more massive HCG~57A, in which star formation is suppressed by at least a factor of 25. \citet{Alataloetal15} argue that the significant quantities of cold molecular gas in HCG~57A must be turbulent to explain this high suppression factor, and that this could be explained if a shock ring were still expanding through the galaxy, injecting turbulence, while the equivalent shocks in HCG~57D have already passed through the entire galaxy disk. 

Although HCG~57 is one of the most \Hi\ deficient compact groups in the sample of \citet{VerdesMontenegroetal01}, Very Large Array observations trace a diffuse cloud of neutral hydrogen extending across the HCG~57A/C/D triplet, peaking on HCG~57D and with a velocity distribution that overlaps the velocities of galaxies C and D but not HCG~57A \citep{Jonesetal23}. H$\alpha$ emission is observed from the star-forming ring of HCG~57D, with a faint diffuse structure connecting it to the center of HCG~57A \citep{TorresFloresetal14}. In the X-ray, \textit{ROSAT} and \textit{ASCA} observations detected emission from a $\sim$1~keV diffuse IGrM centred on the triplet \citep{Mulchaeyetal96,Ponmanetal96,Fukuzawaetal02} and extending to at least 8\arcm\ radius \citep{Mulchaeyetal03}. Table~\ref{tab:veloc} lists the morphological type and recession velocity of the member galaxies, drawn from \citet{Hicksonetal89} and \citet{Hicksonetal92}. 

\begin{deluxetable}{lccccc}
\tablewidth{0pt}
\tablecaption{\label{tab:veloc}Properties of group member galaxies drawn from \citet{Hicksonetal89} and \citet{Hicksonetal92}}
\tablehead{
\multicolumn{2}{c}{Galaxy} & \colhead{RA} & \colhead{Dec.} &  \colhead{Type} & \colhead{$cz$}\\
\colhead{HCG~57} & \colhead{NGC} & \colhead{(deg.)} & \colhead{(deg.)} & & \colhead{(\kmps)} 
}
\startdata
A & 3753 & 174.47409 & 21.98084 & Sb  & 8727 \\
B & 3746 & 174.43198 & 22.00933 & SBb & 9022 \\
C & 3750 & 174.46564 & 21.97384 & E3  & 9081 \\
D & 3754 & 174.47966 & 21.98564 & SBc & 8977 \\
E & 3748 & 174.45486 & 22.02576 & S0a & 8992 \\
F & 3751 & 174.47535 & 21.93609 & E4  & 9594 \\
G & 3745 & 174.43585 & 22.02083 & SB0 & 9416 \\
H & - & 174.46122 & 22.01187 & SBb & 9042 \\
\enddata
\end{deluxetable}

For convenience of comparison with \citet{Alataloetal14} we adopt a distance of 132~Mpc for the group, making 1\arcs\ equivalent to 640~pc. We note that for HCG~57C and F a  redshift-independent distance of 138~Mpc ($\pm$15\%) has been estimated based on the fundamental plane of elliptical galaxies \citep{Saulderetal13,Saulderetal16}. This is consistent within the uncertainty with our adopted distance. Velocities are heliocentric and uncertainties are given at the 1$\sigma$ significance level unless otherwise stated.

\section{Observations and Data Reduction}

\subsection{X-ray observations}
\label{sec:Xray}

HCG~57 has been observed by both \chandra\ and \xmm, and we describe these observations and their reduction below. Table~\ref{tab:Xray_data} lists the ObsIDs, dates and cleaned exposure times of each observation. Throughout the X-ray analysis we adopted a Galactic hydrogen column of 2.35$\times$10$^{20}$~cm$^{-2}$, drawn from the HI4PI survey \citep{HI4PI16}. We also adopted the solar abundance tables of \citet{GrevesseSauval98}, and spectral fitting was performed using \textsc{Xspec} v12.11.0m \citep{Arnaud96}.

\begin{deluxetable}{lcc}
\tablewidth{0pt}
\tablecaption{\label{tab:Xray_data}Summary of the X-ray observations}
\tablehead{
\colhead{ObsID} & \colhead{Observation Date} & \colhead{Exposure (ks)}
}
    \startdata
    \multicolumn{3}{l}{\textit{Chandra}}\\
    20583 & 2019 Mar 05 & 34.6 \\
    20584 & 2019 Feb 09 & 14.9 \\
    22055 & 2019 Feb 19 & 23.4 \\
    22120 & 2019 Feb 22 & 39.7 \\
    22121 & 2019 Feb 23 & 40.0 \\
    22122 & 2019 Feb 25 & 15.0 \\
    22138 & 2019 Mar 06 & 27.5 \\
    \multicolumn{2}{r}{Total:} & 195.1\\
    \multicolumn{3}{l}{\textit{XMM-Newton}}\\
    0206090101 & 2004 Dec 21 & 22.2 \\
    0206090201 & 2005 May 20 & 7.1 \\
    \multicolumn{2}{r}{Total:} & 29.3\\
    \enddata
\end{deluxetable}

\subsubsection{Chandra}
HCG~57 was observed by \chandra\ for a total of $\sim$195~ks during cycle~19, in seven exposures. These observations are contained in the \chandra\ Data Collection (CDC) 319 \dataset[DOI: 10.25574/cdc.319]{https://doi.org/10.25574/cdc.319}. The ACIS-S3 CCD was at the focal point, operating in VF mode. A summary of the \chandra\ mission is provided in \citet{Weisskopfetal02}. We reduced the resulting data sets following the \chandra\ analysis threads\footnote{http://cxc.harvard.edu/ciao/threads/index.html} and the approach described in \citet{OSullivanetal17}, using \textsc{ciao} 4.13 \citep{Fruscioneetal06} with CALDB 4.9.4. Periods of high flaring background were identified using the \textsc{lc\_clean} script applied to the combined lightcurve of the seven observations, and removed. All observations were then projected onto a common tangent point, and we created combined images and exposure maps. The 0.5-2~keV band was used to examine the spectrally soft diffuse emission.

Point sources were identified from a combined 0.5-7~keV image using the \textsc{wavdetect}. Combined exposure and point spread function (PSF) maps were used, with the individual PSF maps weighted by their associated exposure maps, and using a 90\% encircled energy fraction. False detections associated with gas structures (i.e., cases where \textsc{wavdetect} flagged a clearly extended clump of emission as a point source) were identified by visual inspection and removed from the source list. Point sources in the resulting "cleaned" source list were generally excluded from later analysis of diffuse emission.

Spectra were extracted and response matrices created using the \textsc{specextract} task. Spectra were extracted separately from the event files for each individual observation, then combined using the \textsc{combine\_spectra} script. This reduces complication when fitting, and testing shows that for our observations, taken over a period of $\sim$1 month and using the same CCD, there is no significant difference in results between fitting the combined spectrum or simultaneously fitting the individual spectra from the seven exposures. Background spectra were either extracted from source-free regions on the same CCD, or from the standard blank-sky background files, scaled to match the 9.5-12~keV count rate of the CCD in each observation.

\subsubsection{XMM-Newton}
HCG~57 was observed twice by \xmm, for a total of $\sim$29~ks. The EPIC-pn and MOS instruments operated in full frame mode, with the medium filter. We reduced and analysed the resulting datasets using the \xmms\ Science Analysis Software (SAS) v19.1 \citep{Gabrieletal04} following the approach described in \citet{OSullivanetal17}. 

For an initial analysis, a three-band background filter was applied to identify and remove any periods of high (flaring) background. Point sources were identified using the \textsc{edetect\_chain} task, and where necessary regions corresponding to the 85\% encircled energy radius were excluded. Multi-band images and exposure maps were created for each camera, with examination of soft emission usually using the 0.5-2~keV band, and for point sources the 0.5-7~keV band.

To allow a more detailed analysis of the faint diffuse emission from the group halo, ObSID 0206090101 was also processed using the \xmms\ Extended Source Analysis Software \citep[ESAS,][]{Snowdenetal04}. Initial processing showed MOS~1 CCD~4 and MOS~2 CCD~5 to be in potentially anomalous states, and they were excluded. After cleaning, point sources were identified using the \textsc{cheese} task and masked from all further analysis, with the exception of a source coincident with the core of HCG~57A.

\subsection{Optical spectroscopic observations with GCMS}
\begin{figure}
\includegraphics[width=0.5\textwidth]{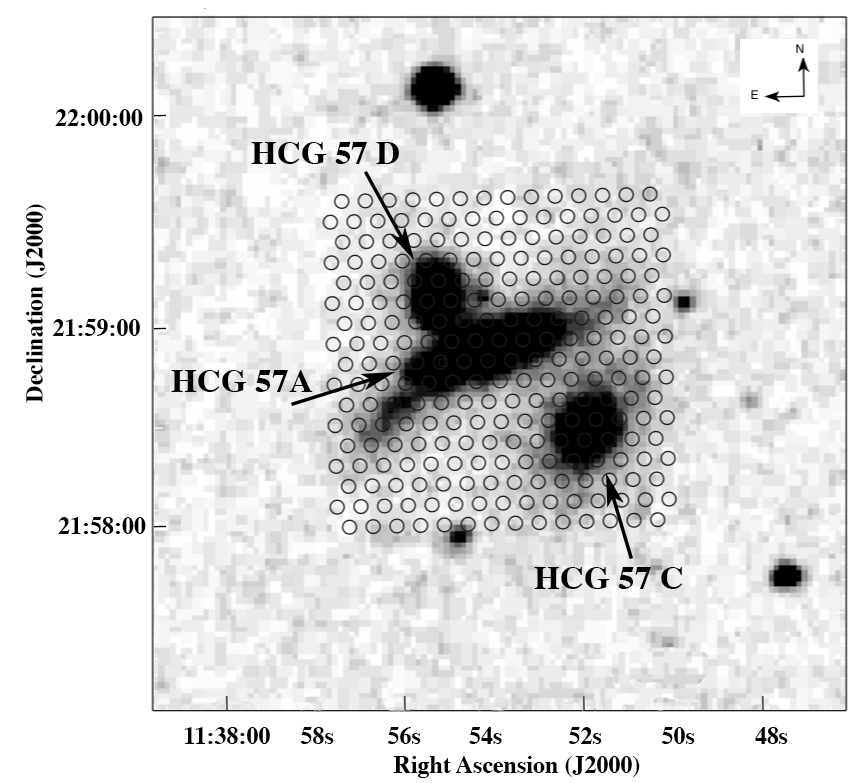}
\caption{Footprint of the fibers from the McDonald Observatory GCMS IFU system for a single pointing superimposed on a high-contrast (Palomar POSSII-g grey scale) image of the inner HCG~57 group. The field of view is 2.8 square acrmins. A six dithers observing pattern was used with to fill in the spaces between the fibers.} 
\label{fig:virusp_foot}
\end{figure}

Integral Field Unit (IFU) spectroscopy was carried out using the George and Cynthia Mitchel Spectrograph (GCMS; \citealt{Hill2008,Blanc2010}), also known as Virus-P, on the 2.7 m Harlan J. Smith telescope at the McDonald Observatory. Data was taken with the VP4 (red grating; spectral resolution 1.5\AA) and VP2 (blue grating; spectral resolution 1.6\AA) on the nights of 2012 Jan 29 and 30th respectively, under conditions of moderate seeing (2 arcsec). The VP4 and VP2 gratings cover the wavelength range of 6200-6850\AA~ and 4700-5350\AA~respectively. GCMS is a 246-fiber system in which six separate dithers are taken at each pointing to completely fill an area of 2.8 square arcminutes. The individual fibers are large (4.16 arcsec diameter on the sky), making the system very sensitive to faint emission. Figure~\ref{fig:virusp_foot} shows the orientation and fiber positions for a single pointing on HCG~57. A total of 1200 s of integration per dither was made in the red, and 2 $\times$ 1200 s in the blue grating, at each dither position.  

The analysis method, including the flux calibration and data cube construction is identical to that described in \citet{Joshi2019}, and will only be summarized briefly here.  The data was processed through the VACCINE data reduction package \citep{Adams2011}. A subset of fibers, selected to be free of emission, were used to perform sky subtraction. Astrometry of each pointing was aligned using Sloan Digital Sky Survey (SDSS; \citealt{York2000}) images through a cross-correlation method  \citep[see][]{Joshi2019}, resulting in the creation of a final flux calibrated 3D spectral cube interpolated onto a cube with 2 $\times$ 2 arcsec$^2$ pixels. The process was performed on data obtained in the blue and red grating setups to form a blue and red cubes.

The analysis of the final data cubes used the LZIFU software toolkit \citep{Ho2016} along with custom written \texttt{python} code. The software fits the continuum emission using a custom version of the PPXF code \citep{Cappellari2004} using model templates. This allows a best fit to the shape of any absorption features, if present. The software then fits any emission-lines present after removal of the continuum.

\section{Results} \label{sec:results}

\subsection{The diffuse intra-group medium}
\begin{figure}
\includegraphics[width=\columnwidth,bb=36 126 577 667]{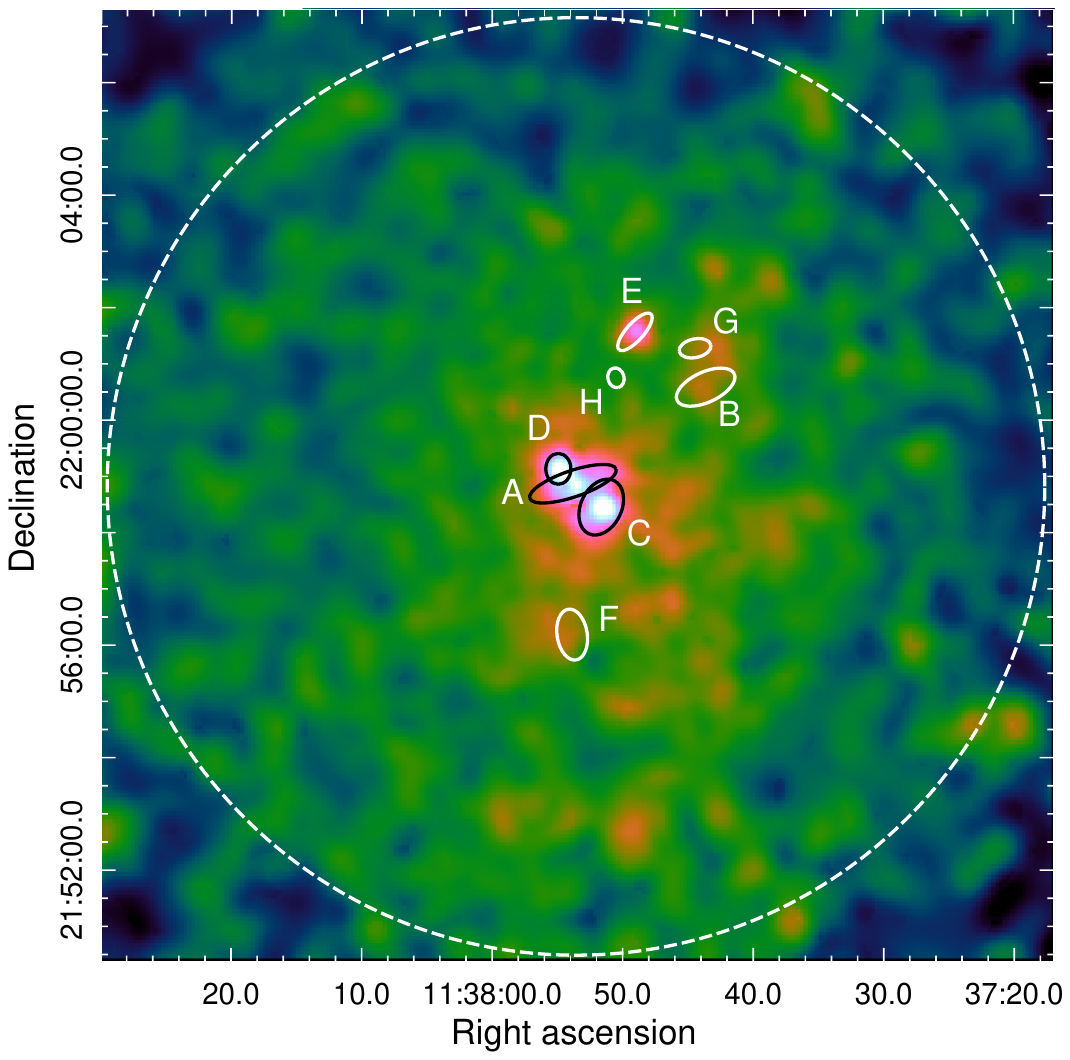}
\caption{\label{fig:XMMim}\xmm\ 0.3-2~keV image of HCG~57, adaptively smoothed with a signal-to-noise ratio of 10, and with point sources outside the member galaxies removed. The \Dtf\ ellipses of the member galaxies are labeled, and the dashed circle marks the 500\arcs\ (320~kpc) radius region within which the IGrM is detected.}
\end{figure}

The extended diffuse hot IGrM detected by \textit{ROSAT} and \textit{ASCA} is visible in the \xmm\ data. Figure~\ref{fig:XMMim} shows a 0.3-2~keV adaptively smoothed image, from which point sources outside the \Dtf\ ellipses of the member galaxies have been removed. As expected, the IGrM is centred on the HCG~57A/C/D triplet. X-ray brightness peaks are visible coincident with the optical centroids of each of these three galaxies, with HCG~57A and D linked by more extended emission, while HCG~57C appears more clearly separated from HCG~57A (see also Figure~\ref{fig:3color}). The separation between the galaxies ($\sim$22\arcs\ between the nuclei of HCG~57A and D) is comparable to the \xmms\ point-spread function and we therefore defer examination of the galaxies to the \chandra\ analysis.

Using the \xmms\ ESAS software, we extracted spectra from a series of concentric annuli centred on HCG~57A. Particle background spectra were also extracted for each region and detector. All spectra were then fitted simultaneously in \textsc{Xspec}. Diffuse emission from the group halo was modelled by a deprojected absorbed APEC thermal plasma model, with an intrinsically-absorbed powerlaw component in the central bin to account for emission from the active galactic nucleus (AGN) in HCG~57A. Residual particle background in each detector was modelled by power laws, and fluorescent lines by Gaussians. The cosmic hard X-ray background was modelled by an absorbed powerlaw with index $\Gamma$=1.46, Galactic and local soft background components were modelled by three APEC models, one un-absorbed with temperature of 0.1~keV, the others absorbed with temperatures 0.1 and 0.25~keV. After initial fitting the temperature of the 0.25~keV component was allowed to fit freely but stayed close to this initial value. Component normalizations were linked between instruments and regions following the approach described in \citet{Snowdenetal04}. 

\begin{figure*}
\centering
\includegraphics[width=0.49\textwidth,bb=30 265 555 770]{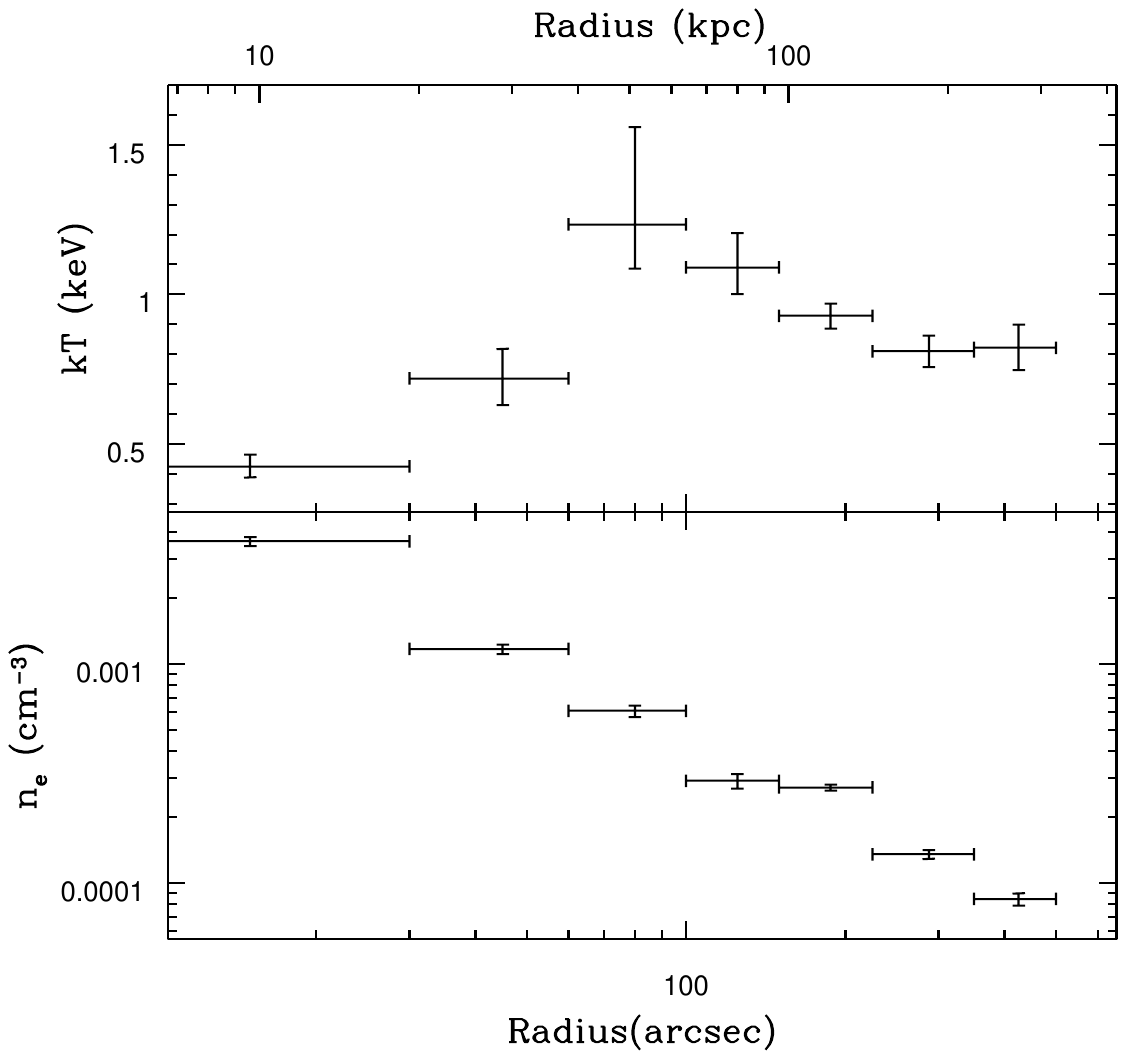}
\includegraphics[width=0.49\textwidth,bb=30 265 555 770]{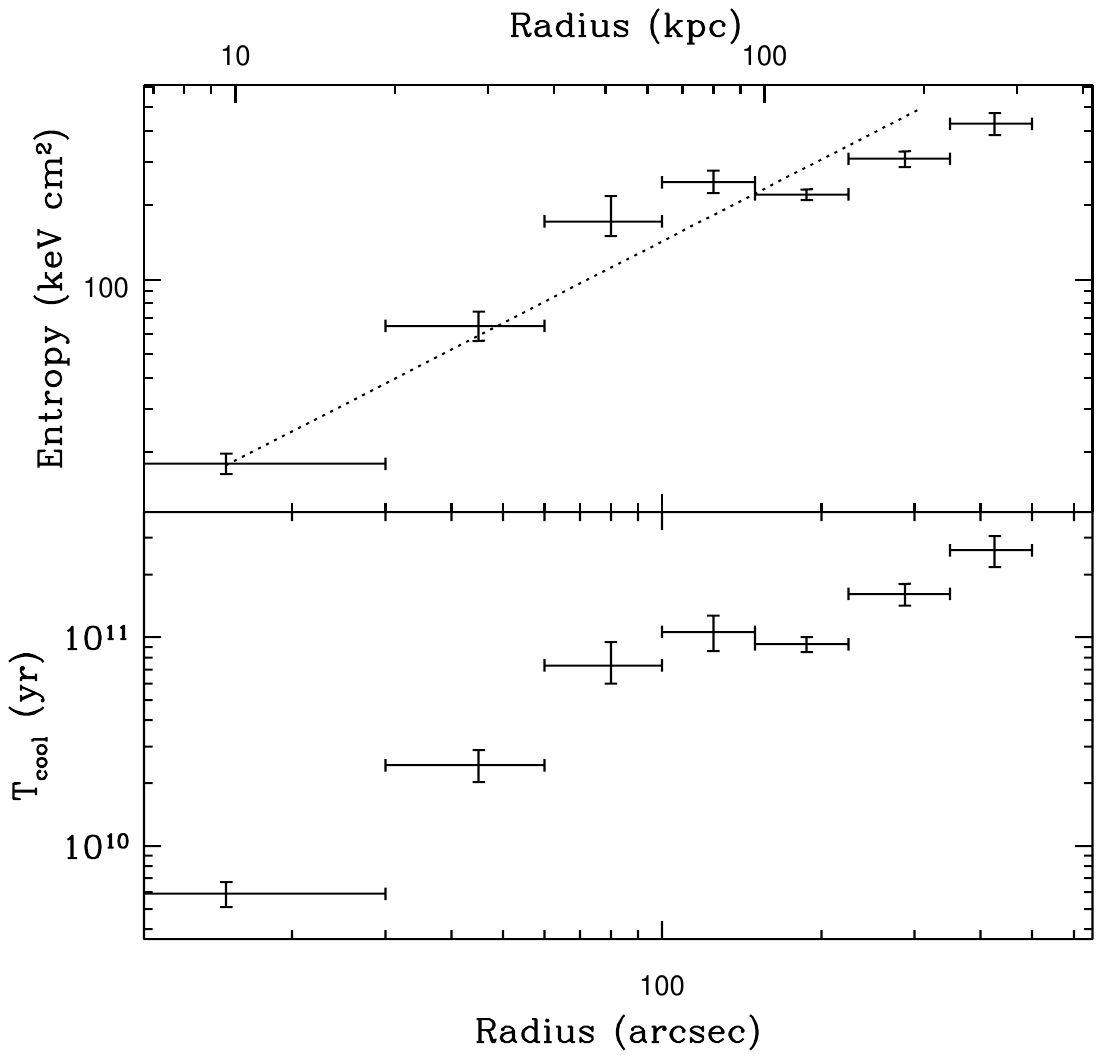}
\caption{\label{fig:XMMprofiles}Radial profiles of gas properties in the hot halo, derived from the \xmms\ deprojection analysis. The dotted line indicates the (arbitrarily scaled) r$^{1.1}$ entropy trend expected for gas that is minimally affected by cooling and non-gravitational heating.}
\end{figure*}

The resulting deprojected radial profiles are shown in Figure~\ref{fig:XMMprofiles}. The previously reported $\sim$1~keV halo is confirmed, with a fairly typical temperature profile for a galaxy group, cool in the centre, peaking at moderate radii and declining in the outskirts, detected to $\sim$500\arcs\ (320~kpc) radius. The abundance of the gas is low, $\sim$0.2\Zsol, but not strongly constrained, and had to be linked between radial bins. Although point sources have been excluded, the temperature in the two central bins will likely be affected by the more complex emission in and around the central galaxies. However, the low central entropy (18.0$\pm$1.7~keV~cm$^2$) and moderate cooling time (5.9$\pm$0.8~Gyr) are not unusual for low-X-ray-luminosity groups \citep[see, e.g.,][]{OSullivanetal17}. The entropy profile has an overall gradient similar to the r$^{1.1}$ behaviour expected for gas which is only weakly affected by cooling and non-gravitational heating processes, consistent with the relatively long cooling times. We find the total pseudo-bolometric luminosity of the IGrM within 500\arcs/320~kpc to be 7.38$\pm$0.22$\times$10$^{41}$~erg~s$^{-1}$.

\subsection{X-ray emission in the central triplet and the HCG~57 A-D bridge}
\begin{figure*}
\includegraphics[width=0.495\textwidth,bb=36 126 577 667]{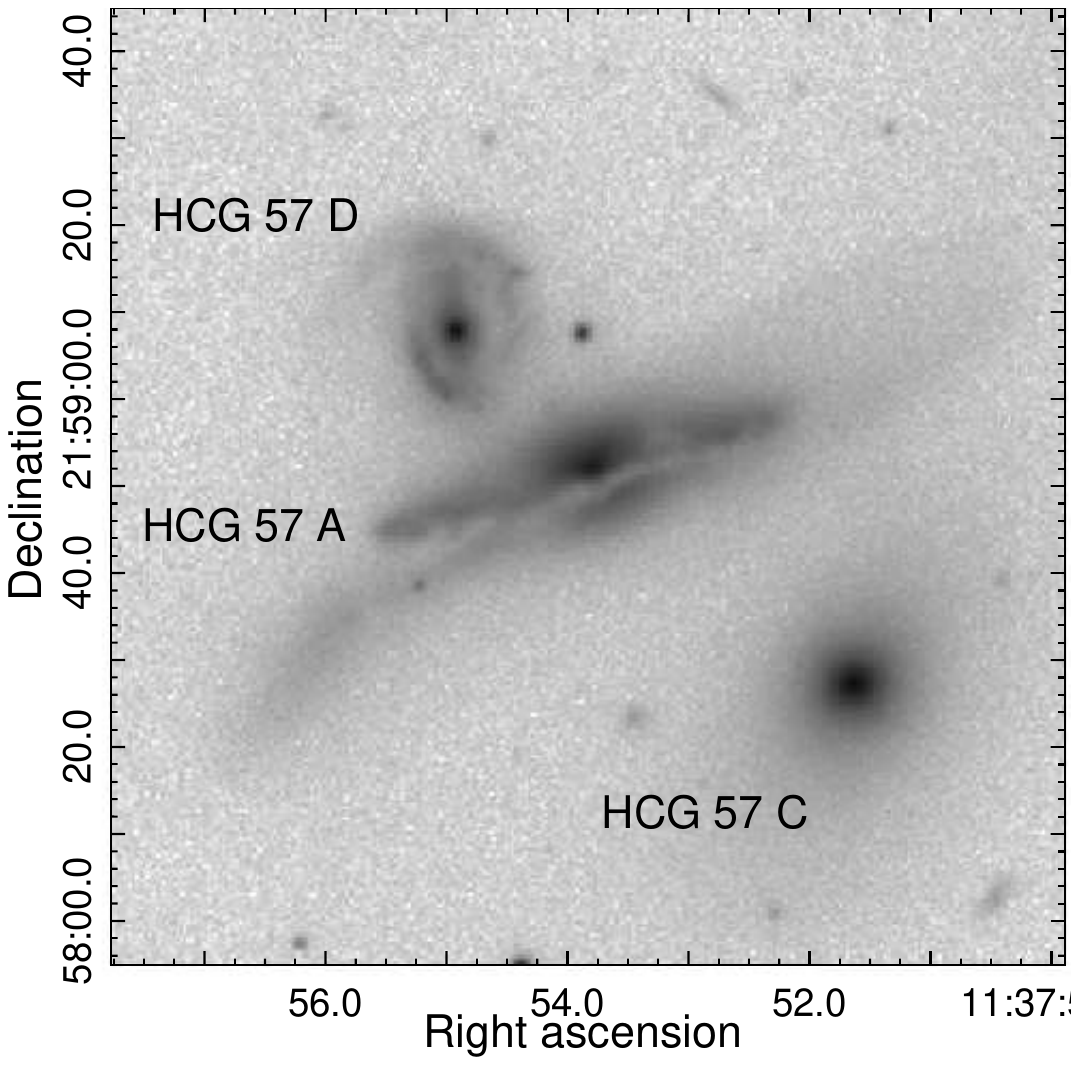}
\includegraphics[width=0.495\textwidth,bb=36 126 577 667]{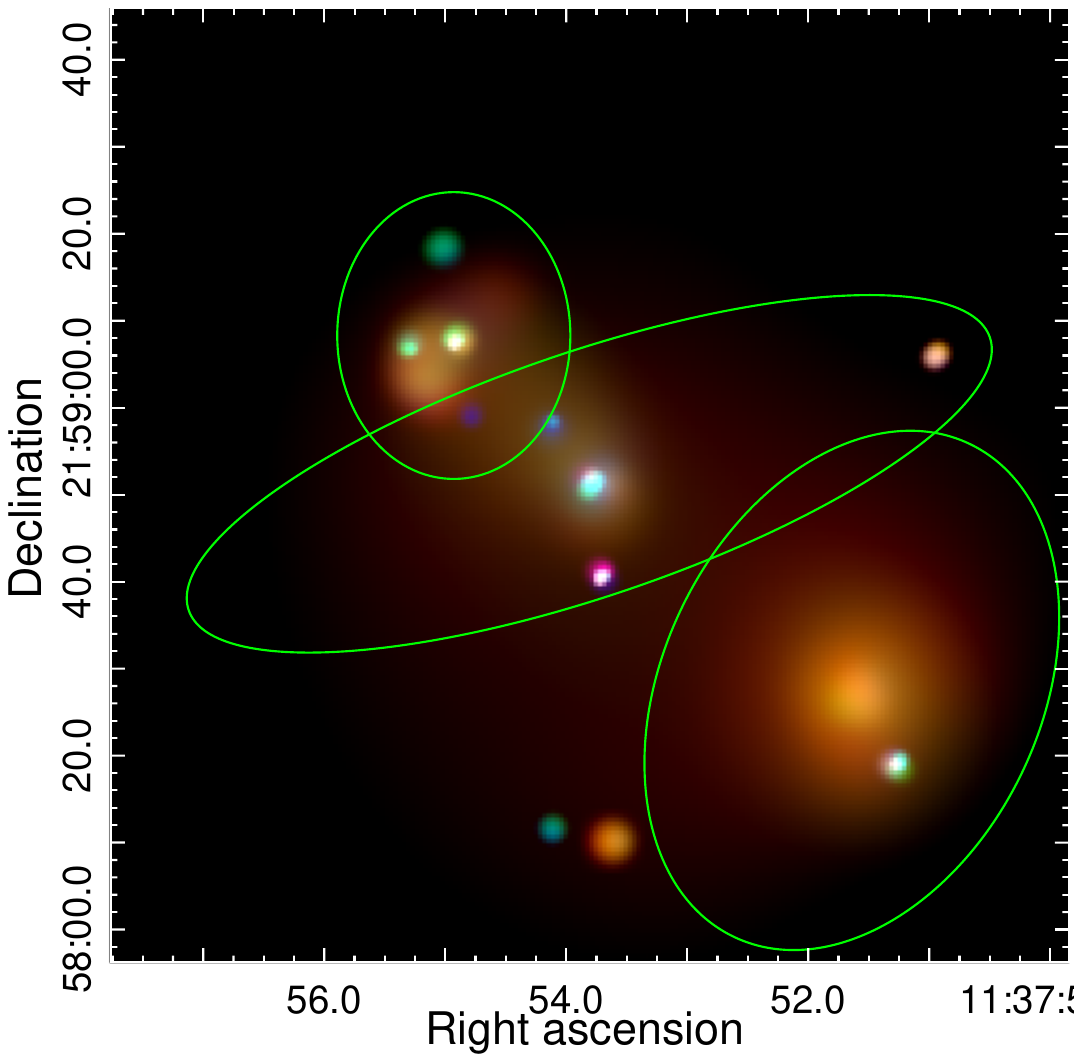}
\caption{\label{fig:3color}SDSS $r$-band and \chandra\ 3-color images of the central triplet, from NE to SW HCG~57 D, A and C. \chandra\ images in the 0.5-1, 1-2, and 2-7~keV bands were used for the red, green and blue channels of the 3-color image, which is adaptively smoothed with an overall signal-to-noise ratio of 3-5. Ellipses mark the approximate \Dtf\ contours of the galaxies.}
\end{figure*}

\begin{figure}
\centering
\includegraphics[width=\columnwidth,bb=50 145 570 660]{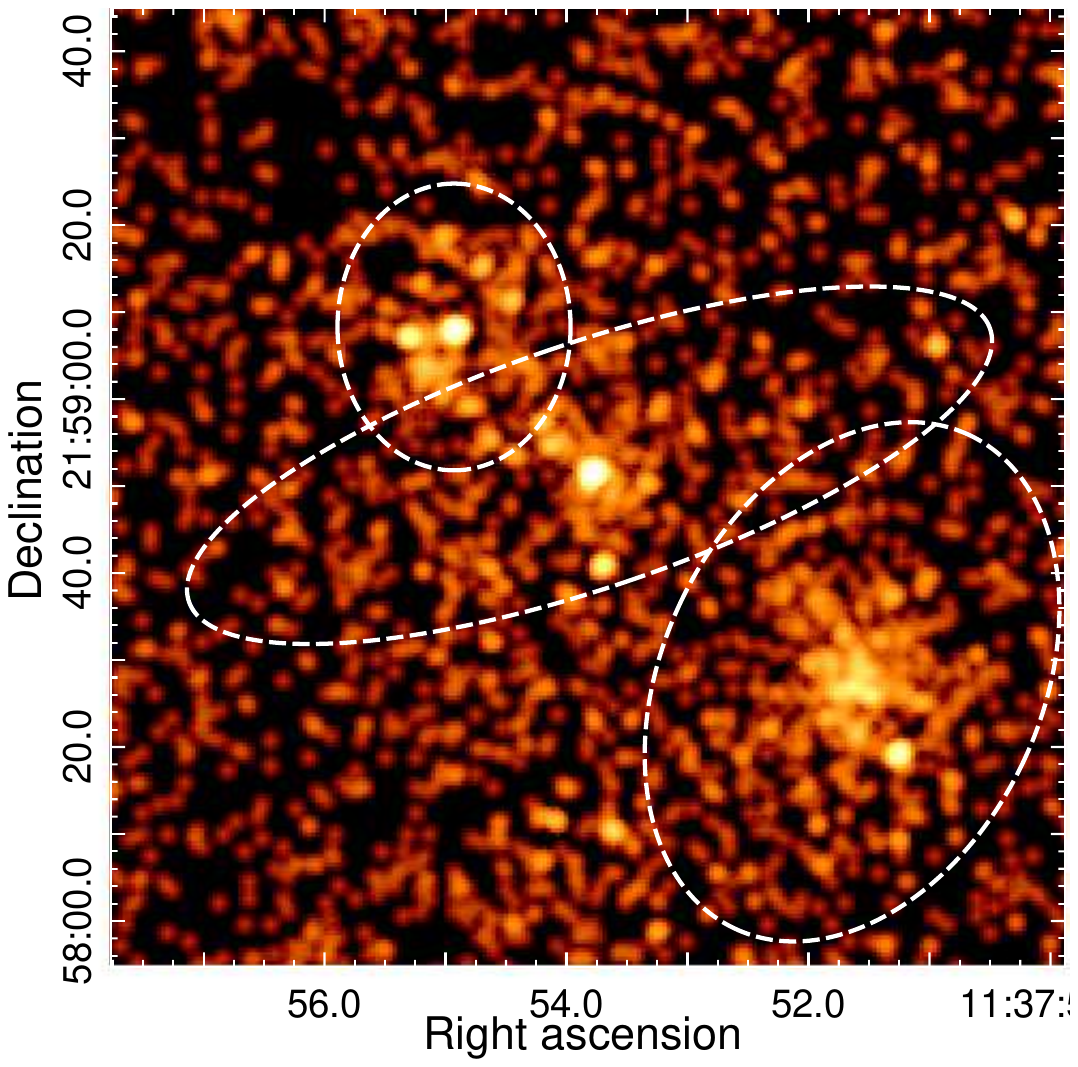}
\caption{\chandra\ 0.5-2~keV image of the central triplet, smoothed with a 1.5\arcs\ Gaussian, with the \Dtf\ ellipses of the galaxies marked. Note that the point source visible inside the HCG~57C \Dtf\ ellipse is offset from the galaxy nucleus by $\sim$10\arcs\ (6.4~kpc).}
\label{fig:HCG57ACD}
\end{figure}

Figure~\ref{fig:3color} shows the HCG~57A/C/D triplet in more detail. The warped disk and dust lane of HCG~57A, and the partial ring of HCG~57D are visible in the SDSS optical image. The \chandra\ adaptively smoothed 3-color image shows several point sources in and around the galaxies, most of which are spectrally hard (blue or white). Spectrally softer (redder) diffuse emission is also visible, e.g., in the elliptical HCG~57C, which appears to have its own relatively relaxed halo. However, in HCG~57A and D, we see diffuse emission which links the cores of the two galaxies, as well as a brighter blob of emission in the SE part of the HCG~57D disk. Figure~\ref{fig:HCG57ACD} shows the \chandra\ 0.5-2~keV image at higher resolution, with a lighter 1.5\arcs\ Gaussian smoothing. This reveals the diffuse emission in the SE quadrant of HCG~57D to be an arc collocated with the brightest part of the partial ring visible in the optical, suggesting it may be related to star formation. However, the emission linking HCG~57A and D appears to be a genuinely diffuse feature. We will focus on this bridge-like structure for the rest of this section.

\begin{figure}
\includegraphics[width=\columnwidth,bb=145 360 390 660]{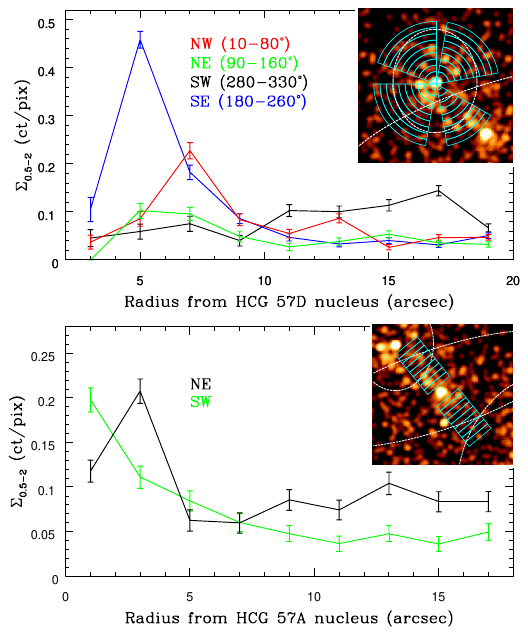}
\caption{\label{fig:SBprofs}\chandra\ 0.5-2~keV surface brightness profiles extracted from the regions shown in the inset panels. The upper panel shows profiles extracted from partial annuli centred on HCG~57D, and the lower panel those extracted using rectangular regions on either side of the disk of HCG~57A. both sets of regions have widths of 2\arcs, and point sources were removed before extraction. The inset panels show the regions overlaid on the 0.5-2~keV \chandra\ image used in Figure~\ref{fig:HCG57ACD}, with \Dtf\ ellipses marked by white dashed lines.}
\end{figure}

To test whether there is, as it appears, an excess of X-ray emission between HCG~57A and D, we extracted surface brightness profiles between the two galaxies, and in other nearby regions. We selected partial annuli in four sectors centred on the nucleus of HCG~57D, with widths of 2\arcs\ and extending to 20\arcs\ radius, so that the profiles in all four sectors extend beyond the ring of enhanced star formation in the galaxy's disk. The regions and resulting profiles are shown in the upper panel of Figure~\ref{fig:SBprofs}. The south-east (SE) and north-west (NW) profiles show strong peaks at 5-7\arcs\ where they cross the ring, but fall off to low values at large radii, as does the north-east (NE) profile. By contrast, the south-west (SW) profile is enhanced beyond $\sim$10\arcs, indicating higher soft X-ray surface brightness outside the ring.

To test whether this enhancement could be caused by a hot gaseous halo associated with HCG~57A, we also extracted profiles to the NE and SW of its edge-on disk. The NE profile covers the region of enhanced emission between the two galaxies. We find that both profiles (Figure~\ref{fig:SBprofs}, lower panel) show stronger X-ray emission toward the centre of HCG~57A, but while the profile to the SW falls off to a background level outside ~10\arcs, the NE profile remains significantly enhanced. This confirms that the X-ray emission between the two galaxies is enhanced compared to their surroundings.

We also extracted spectra on either side of the HCG~57A disk, from 10\arcs$\times$15\arcs\ regions covering a similar area to those used for the surface brightness profiles. The results of fitting absorbed APEC model to the spectrum of the region on the NE side of the galaxy (the A-D bridge) are shown in Table~\ref{tab:diffuse}, with the gas found to have a temperature $\sim$1~keV. A powerlaw can be fitted to the spectrum, but has an unphysically steep index. The equivalent fit for the region on the SW side of the disk is not well constrained (forcing us to fix abundance) but finds a similar temperature (1.03$^{+0.30}_{-0.41}$~keV). The flux in the region is roughly 60\% of that observed in the bridge region. On the basis of these results, we conclude that there is a bridge of hot, X-ray emitting gas linking the cores of the two galaxies.

We cannot know the volumes from which the emission arises, but adopting the simplest geometry, cylinders of radius 5\arcs\ (3.2~kpc) and 15\arcs\ (9.6~kpc) long, we find the electron number density (n$_e$) of the hot gas in the bridge region to be 1.12$^{+0.40}_{-0.36}\times10^{-2}$~cm$^{-3}$, and the gas mass to be 1.00$^{+0.36}_{-0.32}\times10^8$\Msol. The cooling time of the gas is long, 4.01$^{+2.33}_{-2.01}$~Gyr.

\subsection{Ionized gas in HCG~57A and D}
\label{sec:ionized}

\begin{figure*}
    \centering
    \includegraphics[width=0.9\textwidth]{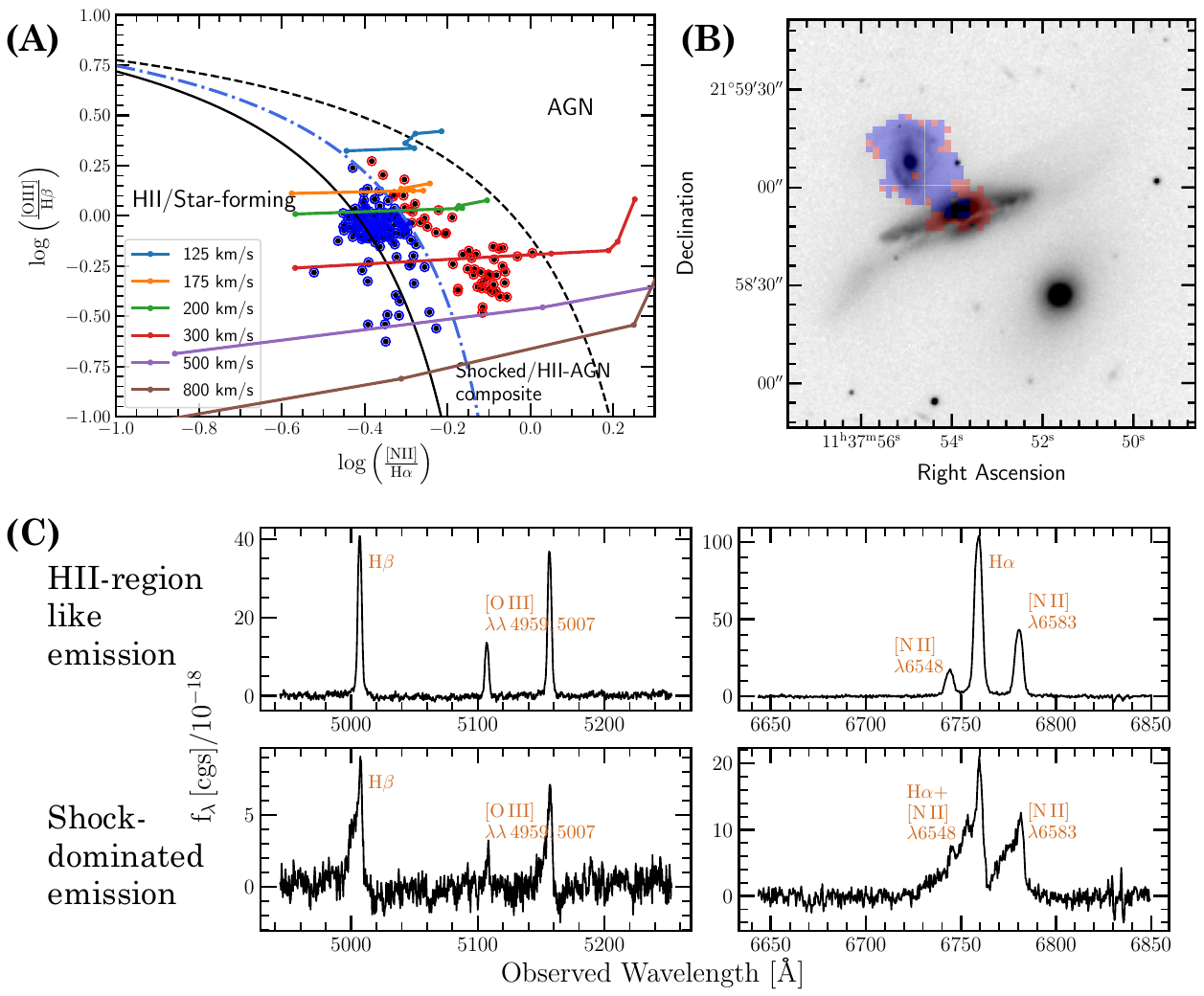}
    \caption{The spatial and spectral differences between shocked and \ion{H}{2}-region-like stellar ionized gas emission. Panel (A): The \ion{N}{2} line diagnostic diagram shows the line ratios computed from the IFU data with shock models of different velocities overlaid \citep{AllenMappings}. The regions are based on the definitions provided by \citet{Kewley06}. The dash-dot line demarcates the shocked (red) vs. stellar (blue) areas of the plot. Panel (B): shocked and stellar spaxels colored according to the spaxels identified in the BPT diagram in panel (A). The underlying image is an SDSS $i$-band image. Panel (C): average spectra corresponding to the spaxels shown in panel (A) with the stellar emission in the top row (blue spaxels) and the shocked emission in the bottom row (red spaxels). The asymmetric line profiles in the shock-dominated emission suggest emission from gas as a range of velocities.}
    \label{fig:shock_vs_stellar_spectra}
\end{figure*}

Emission lines from the optical IFU data offer further clarity on the nature of the ionized gas bridge between HCG~57A and HCG~57D. We fit continuum subtracted pure emission line spectra to extract line fluxes from H$\alpha$, H$\beta$, [\ion{N}{2}], and [\ion{O}{3}], and construct line diagnostic diagrams using these line fluxes \citep[][typically referred to as BPT or VO diagrams]{BPT, Veilleux1987}. Our line diagnostic diagram is shown in panel (A) of Figure~\ref{fig:shock_vs_stellar_spectra}, along with regions corresponding to stellar/\ion{H}{2} regions, stellar+AGN composite, and AGN emission \citep{Kewley06}, and overlaid with the shock models of \citet{AllenMappings}. Prior work by \citet{Rich2011, Rich2014, Rich2015} has shown that shock excited line emission can mimic emission from \ion{H}{2} regions + AGN, appearing in the composite area of the [\ion{N}{2}] line diagnostic diagram. It can be observed that many spaxels from the IFU data for HCG~57 display such composite/shocked emission. Stellar vs. shocked emission is classified through the line diagnostic diagram -- we consider all points to the right of the average line between the \ion{H}{2} and AGN classification lines to be shock excited. Panel (B) of Figure~\ref{fig:shock_vs_stellar_spectra} shows the spatial distribution of spaxels exhibiting stellar and shocked emission. Shocked emission is found in the disk of HCG~57A, at the edges of the gas bridge, and in the outskirts of HCG~57D, while the disk of HCG~57D and the center of the bridge show \ion{H}{2}-region-like stellar spectra. Clear spectral differences between the emission lines from the shocked and stellar spaxels can be observed (panel (C) Figure~\ref{fig:shock_vs_stellar_spectra}). The shocked emission lines are broader and significantly asymmetric relative to the stellar lines. We note that in regions of strong star formation (as in the ring and inner disk of HCG~57D) the stronger \ion{H}{2}-region-like emission may overwhelm any contribution from a tenuous shocked gas component \citep[as in, e.g.,][]{Rich2015}. It is therefore possible that shocked gas extends through much of the disk of HCG~57D, even though it is only clearly detected in the outskirts.

\begin{figure*}
    \centering
    \includegraphics[width=0.9\textwidth]{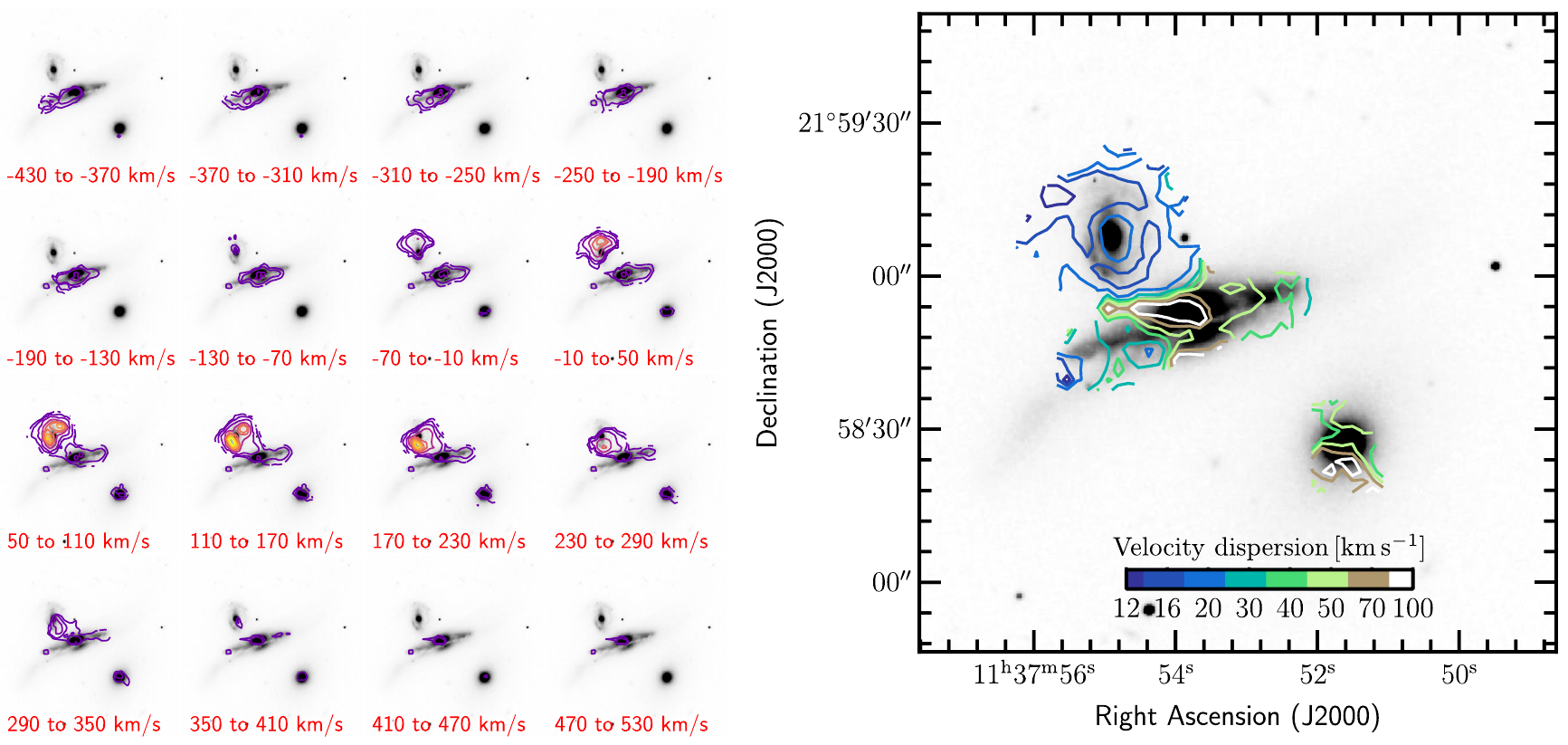}
    \caption{Velocity channel map and velocity dispersion map from H$\alpha$. The channel map shows H$\alpha$ flux contours within each channel relative to systemic velocity. Both maps are overlaid on an SDSS $i$-band image. A systemic velocity of 8850\kmps, chosen as the approximate mean velocity of HCG~57 A and D, is used as the zero point for the channel map corresponding to an average redshift of $z{=}0.0295$.}
    \label{fig:halpha_channel_vdisp_maps}
\end{figure*}

Figure~\ref{fig:halpha_channel_vdisp_maps} shows the velocity channel map and the velocity dispersion map derived from fitting a Gaussian profile to the H$\alpha$ emission. While a rotation curve can be observed for HCG~57A, the contours for HCG~57D appear more complex. HCG~57D displays H$\alpha$ emission concentrated in two ``lobes'' that show up in multiple channels. The bridge of ionized gas between the two galaxies is also clearly visible appearing in all channels between 170 km/s to 530 km/s. Along the ionized gas bridge the velocity dispersion can also be observed to change rapidly with significant velocity dispersion toward HCG~57A.

\subsection{Diffuse X-ray emission in member galaxies}
Diffuse (or unresolved) X-ray emission is visible in several of the member galaxies in the \chandra\ data. To examine the this emission, we extracted \chandra\ spectra from within their \Dtf\ ellipses (and excluding any point sources) combined the spectra and responses, and fitted them in \textsc{Xspec}. For HCG~57A, C and D, where the \Dtf\ ellipses overlap (see Figure~\ref{fig:HCG57ACD}) we exclude the overlap region between A and D when extracting the spectra for A, and the overlap region between A and C when extracting the spectra for C. In practice, only the bridge region linking HCG~57A and D contains a significant part of the diffuse emission in the two galaxies.

Spectra were fitted with absorbed APEC thermal plasma or power law models, or a combination of the two. The emission is generally faint and in many cases it was not possible to clearly constrain the abundance of the thermal plasma; in these cases we fixed the abundance at a value close to its best fitting value when free to fit. The results of the fits are shown in Table~\ref{tab:diffuse}. The diffuse emission in HCG~57B, D and G is best described by simple thermal plasma models, while HCG~57F is well-modelled by a $\Gamma$$\simeq$1.1 powerlaw. Based on the typical ratio of X-ray to K-band luminosity in early-type galaxies \citep{Borosonetal11} we would expect a luminosity within a factor of $\sim$2 of $\sim$8$\times$10$^{39}$\ergps; it therefore seems probable that the emission in HCG~57F arises from the X-ray binary population.

\begin{deluxetable*}{llcccccc}
\tablewidth{0pt}
\tablecaption{\label{tab:diffuse}Best-fitting spectral model parameters for diffuse emission in the major galaxies}
\tablehead{
\colhead{Galaxy} & \colhead{Model$^a$} & \colhead{kT} & \colhead{Abund.$^{b}$} & \colhead{L$^{\rm AP}_{0.5-7}$} & \colhead{$\Gamma$} & \colhead{L$^{\rm PL}_{0.5-7}$} & \colhead{$\chi^2$/d.o.f.} \\
& & \colhead{(keV)} & \colhead{(\Zsol)} & \colhead{(10$^{40}$ erg~s$^{-1}$)} & & \colhead{(10$^{40}$ erg~s$^{-1}$)} & 
}
\startdata
HCG~57A & AP    & 1.13$^{+0.17}_{-0.13}$ & 0.14$^{+0.16}_{-0.08}$ & 2.05$^{+0.24}_{-0.25}$ & - & - & 42.5/38 \\
        & AP+PL & 1.08$^{+0.19}_{-0.14}$ & 0.3 & 0.94$^{+0.65}_{-0.71}$ & 1.80$^{+0.84}_{-1.84}$ & 1.81$^{+0.16}_{-0.09}$ & 41.67/37 \\
HCG~57B & AP    & 0.53$^{+0.74}_{-0.26}$ & 0.3 & 0.50$^{+0.38}_{-0.27}$ & - & - & 13.81/10 \\
HCG~57C & AP    & 0.96$\pm$0.05 & 0.29$^{+0.19}_{-0.11}$ & 3.33$^{+0.29}_{-0.28}$ & - & - & 60.49/43 \\
        & AP+PL & 0.89$^{+0.07}_{-0.05}$ & 1.0 & 2.58$^{+0.33}_{-0.34}$ & 1.01$^{+0.43}_{-0.48}$ & 2.32$\pm$+0.54 & 45.76/42\\
HCG~57D & AP    & 1.11$^{+0.15}_{-0.12}$ & 0.3 & 1.01$\pm$0.14 & - & - & 9.96/13 \\
HCG~57F & PL    & - & - & - & 1.08$^{+0.59}_{-0.57}$ & 1.27$^{+0.44}_{-0.39}$ & 25.29/22 \\
HCG~57G & AP    & 0.60$^{+0.67}_{-0.31}$ & 0.3 & 0.90$^{+1.25}_{-0.48}$ & - & - & 0.12/2 \\
\hline
A-D bridge & AP & 1.00$^{+0.38}_{-0.27}$ & 0.13$^{+0.35}_{-0.12}$ & 0.51$^{+0.11}_{-0.11}$ & - & - & 3.56/5 \\
\enddata
\tablecomments{$^a$ Model indicates the spectral model in use, APEC thermal plasma (AP) or power law (PL).\\
$^b$ Where abundances were not constrained, they were fixed and the value is listed without uncertainties.}
\end{deluxetable*}

The spectrum of HCG~57A can be reasonably well described by a $\sim$1.1~keV APEC thermal plasma, or by such a plasma plus a $\Gamma$$\approx$1.8 power law. The APEC+powerlaw model is not a statistically significant improvement on the APEC-only fit, but may capture the emission from the unresolved X-ray binary population of the galaxy.

The E3 galaxy HCG~57C is, in terms of diffuse emission, the X-ray brightest galaxy in the group. Its emission is significantly better described by an APEC+powerlaw model than by an APEC model alone. In either case the APEC has kT$\approx$0.9~keV, but inclusion of the powerlaw pushes the abundance to higher values and leaves it unconstrained; we fix it at solar. The powerlaw component has a relatively shallow slope, $\Gamma\approx$1.0.

The spatial distribution of the diffuse emission in HCG~57C is relatively regular, and we therefore attempted to model it. Using a combined image of the S3 CCDs in the 0.5-2~keV band, we modeled the galaxy emission with a $\beta$-model, plus a flat component for the background, with both components folded through an appropriate exposure map. For a model constrained to be circular, the best-fitting core radius and $\beta$ were 3.33$^{+1.16}_{-0.94}$\arcs\ (2.13$^{+0.74}_{-0.60}$~kpc) and 0.50$^{+0.04}_{-0.03}$ respectively. If the model is not constrained to be circular, its ellipticity and position angle depend on whether the emission associated with HCG~57A and D is excluded from the image, and on the size of the region used to exclude them. However, examination of residual images confirms that the circular model removes the emission from HCG~57C cleanly with no obvious remaining structures, suggesting that its halo is relaxed. 

\subsection{X-ray point sources}
The distance of HCG~57 means that even with our relatively deep \chandra\ data, we are only sensitive to fairly luminous point sources. We used the Portable Interactive Multi-Mission Simulator \citep[PIMMS,][]{Mukai93} to estimate the sensitivity cutoff for the combined \chandra\ observations. Using the 0.5-7~keV band (appropriate for measuring numbers of counts from spectrally hard sources) for an on-axis point source we would expect only a single background count in the area of the point source, so would require 10 source counts for a 3$\sigma$ significance detection. Adopting a powerlaw model with $\Gamma$=1.7 and Galactic absorption, we found that such a source would have flux $F_{0.5-7}\geq 5.87\times 10^{-16}$~erg~cm$^{-2}$~s$^{-1}$. For our adopted distance of 132~Mpc, this is equivalent to $L_{0.5-7}\geq 1.22\times 10^{39}$~erg~s$^{-1}$. In practice, most sources are found some distance from the optical axis, with somewhat larger PSFs and background contributions. For the most compact source we identify on the S3 CCD (where HCG~57A, C and D fall) we would require 11 source counts for a 3$\sigma$ detection, equivalent to $L_{0.5-7}\geq 1.34\times 10^{39}$~erg~s$^{-1}$. For the sources falling on the S2 CCD (HCG~57B, E, G and H), the smallest source would require 15 source counts for a 3$\sigma$ detection, equivalent to a flux limit of $F_{0.5-7}\geq 1.24\times 10^{-15}$~erg~cm$^{-2}$~s$^{-1}$, and $L_{0.5-7}\geq 2.60\times 10^{39}$~erg~s$^{-1}$. Individual X-ray binaries are classified as Ultra-Luminous X-ray sources (ULXs) if their 0.5-8~keV luminosity exceeds 10$^{39}$~erg~s$^{-1}$ \citep{Swartzetal04}. For our assumed spectrum the flux in the 0.5-8~keV is a factor 1.075 greater than in the 0.5-7~keV band. We can therefore say that any source we detect, if it is a single X-ray binary and not a collection of sources, is likely to be a ULX.

Scaling from the cumulative luminosity function of background sources  \citep{Kimetal07a}, we can estimate the number of unrelated background AGN we would expect to see within the \Dtf\ ellipses of the group-member galaxies. For the galaxies falling on S3 (or S2), the summed area of the galaxies is $\sim$1.77 ($\sim$0.91) arcmin$^2$. Based on the sensitivity limits for the two CCDs we find that we would expect to see 1-2 background sources in the \Dtf\ ellipses of those galaxies on S3 and 0-1 sources in those on S2.

\begin{figure}
\centering
\includegraphics[width=\columnwidth,bb=58 113 560 690]{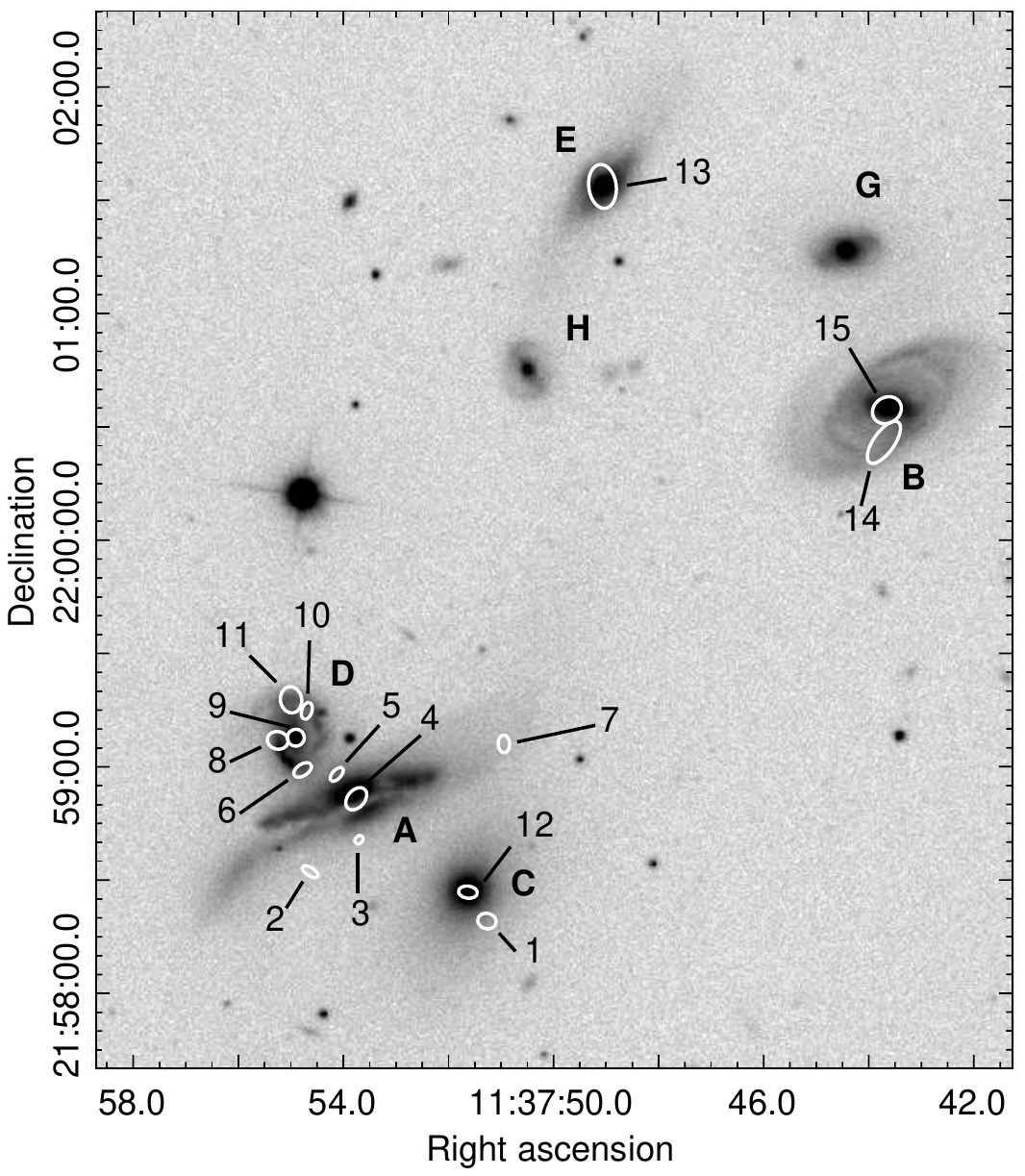}
\caption{\label{fig:PS} SDSS $g$-band image of HCG~57, with point source regions overlaid as white ellipses. Galaxies are labeled and point sources numbered as in Table~\ref{tab:PSsum}}
\end{figure}

Of the point sources identified by \textsc{wavdetect}, fifteen fall within the $D_{25}$ ellipses of group member galaxies. Their positions are marked on Figure~\ref{fig:PS} and their properties are summarized in Table~\ref{tab:PSsum}. Five of these sources correspond to the optical centroids of HCG~57A, B, C, D and E, and may represent their AGN. Prior optical studies have identified all of these except HCG~57C as AGN hosts \citep{Martinezetal10}. The source at the core of HCG~57C is relatively weak and only marginally visible above 2~keV, and is possible that we are simply detecting the peak of the combined emission from hot gas and the stellar population of the galaxy, rather than a genuine point source. As its nature is uncertain, we have conservatively chosen to include it in our analysis. 

Two non-nuclear sources (numbers 3 and 11) are $<$1\arcs\ from \textit{Spitzer} infrared sources. These could be the two unrelated background sources we expect to find within the line of sight of the member galaxies, but since neither infrared source has a measured redshift or is detected at another frequency we cannot say whether such associations may be meaningful. A third X-ray source (number 5), located between HCG~57A and D, has a \textit{Spitzer} infrared source $\sim$1.3\arcs\ from its centroid, within the 90\% X-ray uncertainty ellipse. A faint knot of optical emission is visible at the position of the infrared source in the SDSS $g$-band image. With no further information it is unclear whether this is a background source, a substructure within HCG~57A, or potentially a knot of star formation in the  bridge linking the galaxies.

\begin{deluxetable*}{lccccccccc}
\tablewidth{0pt}
\tablecaption{\label{tab:PSsum}List of point sources and their properties}
\tablehead{
\colhead{Source} & \colhead{R.A.} & \colhead{Dec.} & \colhead{Radii} & \colhead{P.A.} & \colhead{net counts} & \colhead{L$_{0.5-7 keV}$} & \colhead{log(N$_{H,FM}$)} & \colhead{kT$_{FM}$} & \colhead{L$_{0.5-7 keV,FM}$}\\
 & \colhead{(J2000)} & \colhead{(J2000)} & \colhead{(\arcs)} & \colhead{(\degree)} & \colhead{(0.5-7~keV)} & \colhead{(10$^{40}$\ergps)} & \colhead{(cm$^{-2}$)} & \colhead{(keV)}
 & \colhead{(10$^{40}$\ergps)} 
 }
\startdata
\multicolumn{7}{l}{\textit{on S3}} \\
1  & 11 37 51.26 & +21 58 19.1 & 2.49,2.06 & 164.67 & 79.4$\pm$10.1 & 1.14$\pm$0.14 & 22$\pm$0.2 & - & 2.19$^{+0.57}_{-0.80}$\\
2  & 11 37 54.63 & +21 58 32.2 & 2.37,1.09 & 147.54 & 9.9$\pm$4.0 & 0.13$^{+0.06}_{-0.05}$ & 24\tablenotemark{d} & - & $<$34.85\\
3\tablenotemark{a}  & 11 37 53.70 & +21 58 40.6 & 1.31,1.00 &  45.49 & 84.2$\pm$9.5 & 1.27$\pm$0.14 & 22.4$\pm$0.01 & - & 3.48$^{+0.90}_{-0.72}$\\
4  & 11 37 53.75 & +21 58 51.5 & 3.39,2.28 &  49.65 & 339.6$\pm$20.8 & 13.59$^{+4.88}_{-2.81}$ & - & - & - \\
5\tablenotemark{b}  & 11 37 54.12 & +21 58 57.9 & 2.26,1.16 &  47.05 & 33.8$\pm$6.8 & 0.46$^{+0.10}_{-0.09}$ & 22.3$\pm$0.3 & - & 1.39$^{+0.80}_{-0.51}$\\
6  & 11 37 54.77 & +21 58 59.1 & 2.66,1.53 &  33.25 & 17.2$\pm$5.9 & 0.25$\pm$0.08 & $>$21.8 & - & 0.87$^{+54.4}_{-0.53}$ \\
7  & 11 37 50.94 & +21 59 06.0 & 2.35,1.53 &  91.88 & 16.2$\pm$4.9 & 0.23$^{+0.07}_{-0.06}$ & $<$22.4 & - & $<$0.55\\
8  & 11 37 55.25 & +21 59 06.9 & 2.81,2.35 & 168.37 & 30.5$\pm$7.8 & 0.44$\pm$0.10 & $<$21.8 & - & 0.35$^{+0.21}_{-0.07}$\\
9  & 11 37 54.90 & +21 59 07.6 & 2.28,2.25 &  23.93 & 90.6$\pm$10.9 & 1.44$^{+0.18}_{-0.17}$ & - & - & - \\
10 & 11 37 54.69 & +21 59 14.8 & 2.31,1.35 &  71.75 &  9.3 $\pm$4.5 & 0.14$^{+0.06}_{-0.05}$ & $<$22 & - & $<$0.22 \\
11\tablenotemark{c} & 11 37 54.99 & +21 59 17.7 & 3.53,2.82 &  99.84 & 13.6$\pm$6.4 & 0.21$^{+0.08}_{-0.07}$ & $<$21.8 & - & $<$0.35 \\
12 & 11 37 51.62 & +21 58 26.8 & 2.51,1.62 & 173.16 & 26.7$\pm$7.43 & 0.36$^{+0.10}_{-0.09}$ & - & 0.6 & 0.40$^{+0.23}_{-0.24}$\\
\multicolumn{7}{l}{\textit{on S2}} \\
13 & 11 37 49.07 & +22 01 33.66 & 5.85,3.65 & 97.29 & 76.7$\pm$10.3 & 1.33$^{+1.67}_{-1.65}$ & - & $>$1.5 & 1.00$^{+0.26}_{-0.21}$ \\
14 & 11 37 43.71 & +22 00 25.88 & 6.56,2.76 & 54.78 & 36.8$\pm$7.8 & 0.66$^{+0.13}_{-0.12}$ & $<$21.8 & - & 0.87$^{+0.22}_{-0.32}$ \\
15 & 11 37 43.65 & +22 00 34.39 & 4.00,3.41 & 33.09 & 18.0$\pm$6.3 & 0.31$^{+0.10}_{-0.09}$ & $<$21.8 & - & 0.35$^{+0.21}_{-0.12}$ \\
\enddata
\tablenotetext{a}{0.72\arcs\ from SSTSL2~J113753.73+215841.1}
\tablenotemark{b}{1.32\arcs\ from SSTSL2~J113754.08+215859.1}
\tablenotetext{c}{0.9\arcs\ from SSTSL2~J113755.01+215916.8}
\tablenotetext{d}{The parameter is unconstrained, so only the value with highest probability is shown}
\end{deluxetable*}

Two sources are sufficiently luminous to support spectral analysis, numbers 4 and 9, located in the nuclei of HCG~57A and D respectively. We extracted spectra for each source using local background regions, defined as elliptical annuli with the same centroid, axis ratio and position angle as the source ellipse, but with radii 1.5-3$\times$ larger than the source region. For source 4 in HCG~57A we found that the best fitting model was a powerlaw with both Galactic and intrinsic absorption (\textsc{phabs*zphabs*powerlaw}). The powerlaw index was  $\Gamma$=1.78$^{+0.37}_{-0.35}$ and the intrinsic absorbing column was N$_H$=2.47$^{+0.76}_{-0.71}\times$10$^{22}$~cm$^{-2}$. We note that the source is located (in projection) at the edge of the dust lane in HCG~57A, which may explain the intrinsic absorption. For source 9 in HCG~57D, a powerlaw with $\Gamma$=1.65$^{+0.34}_{-0.33}$ and Galactic absorption provided the best fit. The unabsorbed luminosity from these fits is reported in Table~\ref{tab:PSsum}, in the combined column. Adding thermal models to either fit produced either a worse fit or an unconstrained thermal component with flux consistent with zero. 

To measure counts and fluxes where spectral fitting was not possible, we adopted two approaches. In the first, we used the \textsc{srcflux} task. We used the same background regions and adopted a $\Gamma$=1.7 powerlaw model with Galactic absorption to convert counts into flux. The only exception was the HCG~57A nuclear source, where we adopted the best fitting model with intrinsic absorption described above. Table~\ref{tab:PSsum} lists the net 0.5-7~keV counts and luminosity for each source, measured from the combined data. For the HCG~57D nucleus, the flux from spectral fitting is consistent with that estimated by \textsc{srcflux}. The two faintest sources have luminosities $\sim$1.3$\times$10$^{39}$~erg~s$^{-1}$, just above our estimated detection limit. We also used \textsc{srcflux} to examine variability, measuring the flux for each source in each observation where it was detected, or placing a 3$\sigma$ upper limit on flux in observations where sources were not detected at 3$\sigma$ significance. We found that none of the sources showed variability at $>$3$\sigma$ significance.

Our second approach was to use the forward modeling methodology of \citet{Lanzetal22}, which attempts to provide additional constraints on the spectral properties of faint sources by considering hardness ratio as well as count rate. Spectra are simulated with parameters varied across a grid of model parameters, and used to predict the 0.5-8~keV count rate and 2-8/0.5-2~keV hardness ratio. These are compared with the measured values for each source, taking into account the Poissonian uncertainties on the observed numbers of counts in each band. The range of hardness ratios consistent with the observations is determined using the Bayesian Estimation of Hardness Ratios \citep[BEHR;][]{Parketal06}. For each model two parameters are varied, and 1000 simulated spectra are generated for each parameter pair, from which the likelihood of the chosen pair of model parameters producing spectra consistent with the data is determined.

Following the approach of Lanz et al., we use an intrinsically absorbed powerlaw as our baseline model, with fixed index $\Gamma$=1.8, N$_H$ varied between 10$^{21}$ and 10$^{24}$\pcmsq\ in steps of 10$^{0.2}$, and intrinsic 2-10~keV luminosity varied between 10$^{39}$ and 10$^{44}$\ergps, again in steps of 10$^{0.2}$. This model is intended to help us identify sources which are affected by absorption, either from their environment in a group member galaxy or because they are background AGN. We also tested a powerlaw model with no intrinsic absorption (only the Galactic hydrogen column) and an APEC model with fixed solar abundance, kT varied between 0.5 and 2 keV in steps of 0.1~keV, and intrinsic 0.5-7~keV luminosity varied between 10$^{37}$ and 10$^{42}$\ergps\ in steps of 10$^{0.2}$.

The forward modeling approach was applied to all point sources except numbers 4 and 9, where the spectral fits already provide more detailed constraints. The resulting estimates of (or limits on) luminosity, intrinsic absorption column or temperature are reported in Table~\ref{tab:PSsum}. The listed values represent the parameter pair with the highest probability, and uncertainties show the range of values over which the model is not rejected with at least 95\% confidence. In interpreting the results we note that while models can be ruled out (if no parameter pair reproduces the data) a successful model does not necessarily represent the true emission spectrum. Where models are found to be poorly constrained, this may indicate that the spectrum is not well described by the model.

All sources except numbers 10-13 were found to be most consistent with the intrinsically absorbed powerlaw model, though in several cases the data do not strongly constrain the model parameters. Sources 7, 8, 10, 11, 14 and 15 have only upper limits on intrinsic absorption and luminosities consistent with the \textsc{srcflux} estimates. Sources 1, 3, 5 and 6 are all best described by models with absorption above the Galactic level, though for source 6 only a lower limit on absorption is found. As noted above, source 3 may be associated with a \textit{Spitzer} infrared source and the suggestion of intrinsic absorption adds weight to the idea that it may be a background AGN. Source 5 is located between HCG~57A and D, so if it is not a background AGN, intrinsic absorption could imply the presence of dense gas in the bridge between the galaxies. Source 2 is largely unconstrained. The model with the highest intrinsic absorption has the highest probability, leading to a high upper limit on luminosity, but any absorption in the tested range is consistent with the data; we can only say that the APEC and powerlaw model with no intrinsic absorption are disfavoured.

Sources 12 and 13 are found to be inconsistent with powerlaw models, both with and without intrinsic absorption. They are acceptably described by the APEC model, with kT$\simeq$0.6~keV for source 12 and kT$>$1.5~keV for source 13. Source 12 is located at the center of the elliptical HCG~57C, and this finding adds weight to the idea that it may simply be the peak of emission from hot gas in the galaxy rather than a compact source. Source 13 is at the center of the S0a HCG~57E, and the same explanation may apply. We note that the APEC model was also capable of reproducing the properties of sources 10 and 11. These are both faint, with minimal numbers of detected counts, and neither powerlaw or APEC models can be constrained with the current data. We have reported the limits on the absorbed powerlaw model, as this seems more likely to represent these compact, non-central sources.

\section{Discussion}
Previous studies of HCG~57 have provided strong evidence of an off-centre collision between galaxies A and D $\sim$50~Myr ago, with the smaller galaxy having passed through the disk of its larger neighbour \citep{Alataloetal14}. By analogy with other such gas-rich galaxy collisions \citep[e.g., the Taffy galaxies, ][]{Appletonetal15} we might expect to see a bridge of gas drawn out between the two disks. The H$\alpha$ observations of \citet{TorresFloresetal14} provided a hint of such a structure, in the form of weak diffuse emission linking the two galaxies.

Our X-ray and optical line observations provide a clear detection of this bridge, in both warm ionized and hot X-ray emitting gas, linking the cores HCG~57A and D. The 1~keV temperature measured from the bridge emission is similar to the mean temperatures of both galaxies but hotter than the surrounding IGrM emission. Our estimate of the mass of hot gas in the bridge, 1.00$^{+0.36}_{-0.32}$$\times$10$^8$\Msol, is similar to the observed hot gas mass in the bridge of the Taffy galaxies \citep[(0.8-1.3)$\times$10$^8$\Msol,][]{Appletonetal15}. The radiative cooling time of the hot gas is relatively long, $\sim$4~Gyr, though this could be an overestimate if, as in Stephan's Quintet, the bridge contains significant amounts of molecular gas and energy can be radiated away via mid-infrared H$_2$ lines.

\begin{figure*}
\includegraphics[width=0.56\textwidth]{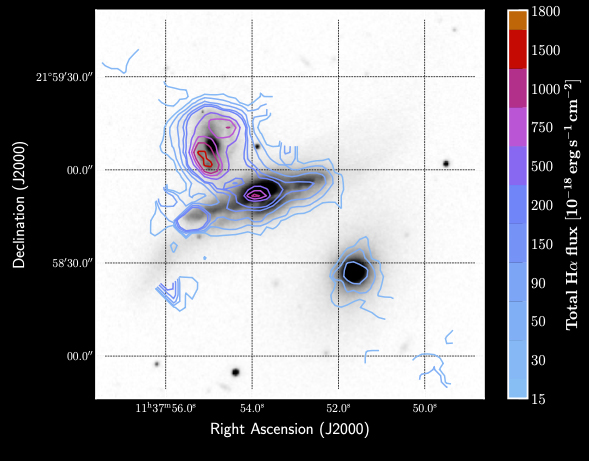}
\includegraphics[width=0.44\textwidth,bb=36 136 560 660]{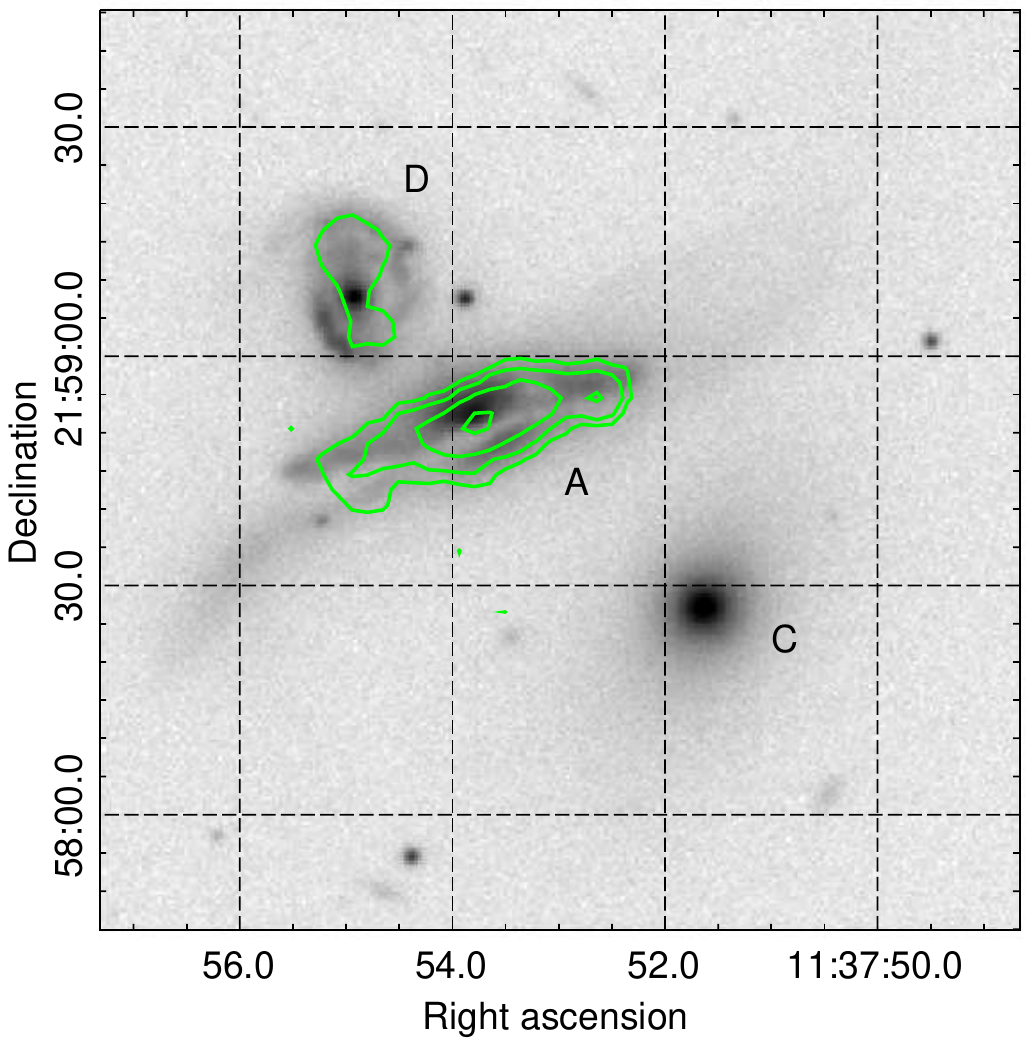}
\caption{\label{fig:HavsCO}Comparison of H$\alpha$ and CO emission in the HCG~57A/C/D triplet. The left panel shows H$\alpha$ contours (4\arcs\ resolution) overlaid on an SDSS $i$-band image, the right panel CARMA CO(1-0) contours (starting at 3$\sigma$ significance and in increasing in steps of 3$\sigma$, 4.6\arcs$\times$3.3\arcs\ HPBW) from \citet{Alataloetal14} overlaid on an SDSS $g$-band image. Note the different distributions of the molecular and ionized gas in the disk of HCG~57D.}
\end{figure*}

Emission from warm, ionized gas (H$\alpha$, H$\beta$, [\ion{O}{3}] and \nii) is detected throughout HCG~57D and along the line of the bridge. Regions with line ratios characteristic of shock heating are visible around the outer edges of HCG~57D, in the disk of HCG~57A, and particularly in the western part of the bridge. Optical line emission from the bridge is seen across a velocity range of $\sim$360\kmps, covering the difference in recession velocities of the two galaxies, and in the regions with shock-like line ratios the line profiles are distinctly asymmetric, with tails extending to shorter wavelengths, indicating emission from gas at a range of velocities.
The overall velocity structure seen in H$\alpha$ is similar to that reported (at lower spatial resolution) in \cii\ by \citet{Alataloetal14}. However, in the disk of HCG~57D the H$\alpha$ emission is strongest in two lobes corresponding to the northwest and southeast sections of the star-forming ring, whereas the \cii\ and CO(1-0) emission is brightest at the northern and southern edges of the disk (Figure~\ref{fig:HavsCO}). This mismatch may indicate the difference between regions still rich in molecular gas and those where it has been depleted by star-formation. The X-ray emission from the disk of HCG~57D is strongest in the southeast sector where H$\alpha$ is also brightest. The peak of the CO emission in HCG~57A also seems to be slightly offset to the southwest of the H$\alpha$ peak. Given the disturbed kinematics of the galaxy disk such an offset is not surprising.

Assuming that the hot gas in the A-D bridge is material shock heated during the galaxy collision, we can estimate the velocity of the interaction based on the shock strength required to heat the precursor gas. As the measured X-ray temperatures in HCG~57A and D are comparable to the temperature in the A-D bridge, it seems likely that the gas in the two galaxies is also heated, by shocks or other processes (e.g., star formation). However, both galaxies are known to contain cold molecular and atomic gas, and this material would inevitably be subject to shocks during a collision. Adopting the strong shock approximation \citep[e.g.,][]{ShullDraine87,Trinchierietal03} for an ideal gas with a ratio of specific heats $\gamma$=5/3, we would expect the post-shock temperature to be:

\begin{equation}
\mathrm{kT} = \frac{3}{16}\mu \mathrm{mv^2 sin^2}\phi,
\end{equation}

\noindent where $\mu$ is the mean molecular weight, m is the mean particle mass, v is the velocity of the shock, and $\phi$ is its obliquity (the angle between the incoming flow and the shock surface). We do not know the obliquity, so adopt $\phi$=90\degree\ for simplicity. For neutral hydrogen with $\mu$=1, we find that a post-shock temperature $kT$=1~keV is expected if v=655\kmps. Note that for an oblique shock with $\phi$$<$90\degree\ the implied velocity would be higher, so our assumption is conservative.

Alternatively, if the pre-shock material were X-ray emitting gas at a temperature of 0.4~keV, similar to that found in the IGrM surrounding the galaxies, the strong shock approximation would no longer hold. In this case, applying the Rankine-Hugoniot jump conditions, we find that a shock of Mach number $\mathcal{M}$=2.3 is required to heat the gas to the observed temperature, equivalent to a velocity of $\sim$750\kmps. 

The distribution of line ratios shown in the  \nii\ line diagnostic diagram in Figure~\ref{fig:shock_vs_stellar_spectra} suggest shocks of velocity $\sim$200-500\kmps, somewhat slower than the estimates from the X-ray. A similar difference in shock velocity estimates is found in the Taffy galaxies and Stephan's Quintet; in the former the X-ray shock velocity is estimated to be 430-570\kmps\ \citep{Appletonetal15} but  optical line diagnostics suggest velocities of 175-200\kmps\ \citep{Joshi2019}, while in the latter the X-ray data are consistent with the 850\kmps\ line of sight infall velocity of the intruder galaxy \citep{OSullivanetal09}, but optical estimates from the ionized gas suggest shock velocities of 200-400\kmps\ \citep{RodriguezBarrasetal14,IglesiasParamoetal12}. 

Simulations of collisions between gas-rich spirals show that their population of gas clouds will experience a range of collision velocities during such interactions \citep[e.g.,][]{YeagerStruck20} depending in part on the effects of rotation in the disks, which can increase or decrease the effective interaction velocity in different parts of the galaxy. The differences in shock velocity estimates may therefore reflect the different apparent collision velocities in different regions of HCG~57A and D. We note that the rotation in the disk of HCG~57A is clearly visible in our H$\alpha$ channel maps (Figure~\ref{fig:halpha_channel_vdisp_maps}) and in the CO and \cii\ maps of \citep{Alataloetal14}. In this context it is notable that the west side of the gas bridge shows the strongest signature of shocks in the ionized gas.

The highly multi-phase nature of the pre-shock interstellar medium (ISM) in both galaxy disks will also play a role. As discussed in detail for Stephan's Quintet  \citep{Guillardetal09} collisions will drive large-scale shocks into the low-density, volume-filling components (e.g., diffuse \Hi, X-ray emitting plasma, etc.) of the ISM in each galaxy. Denser clouds of gas are less likely to undergo collisions with similar clouds in the other galaxy, and the clumpy nature of the ISM means that many dense clouds will be located within larger, medium density structures which will provide some protection from the large-scale shock. However, the rapid increase in the surrounding pressure caused by the shock-heating of the low-density ISM will drive slower shocks into the medium and high-density structures \citep{McKeeCowie75}, and the increased temperature of the ISM may drive increased evaporation in their outer layers \citep{CowieMcKee77}. We can thus expect the initial fast collision shock to produce subsidiary slower shocks in the higher density components of the ISM, leading to a range of apparent shock velocities from different indicators. The HCG~57A/D pair is unusually bright in \cii\ \citep{Alataloetal14} and is one of the strongest emitters of warm, shock-heated H$_2$ in the \textit{Spitzer} sample of \citep{Cluveretal13}. Since the H$_2$ molecule is easily destroyed in strong shocks, it is likely that very slow molecular shocks are also present in the system, and indeed that there is a broad spectrum of shock velocities in different regions and different components of the bridge and ISM. 
The X-ray shock velocity estimates in the Taffy galaxies and Stephan's Quintet both agree with the interaction velocities derived from dynamical arguments, as expected from the arguments above. We therefore expect the greater, X-ray derived velocity estimates for the HCG~57A/D interaction to be closer to the true collision velocity.

The two galaxies are separated by 250\kmps\ in recession velocity, and the northern edge of HCG~57D is $\sim$34\arcs\ ($\sim$22~kpc) from the nucleus of NCG~57A. For collision velocities of 500, 655 or 750\kmps, this implies a timescale for the interaction (i.e., time since the leading edge of HCG~57D passed through HCG~57A) of $\sim$50, 35 or 30~Myr, respectively. This is similar to the timescale of 50~Myr estimated by \citet{Alataloetal14} based on the approximate expansion timescale of the ring-like structure in HCG~57A, and a little greater than the timescale of 25~Myr estimated for the Taffy galaxies \citep{Condonetal93}. Under the conservative assumption that the entire $\sim$10$^8$\Msol\ of X-ray emitting plasma in the bridge was raised to its measured 1~keV temperature by the collision shock, we can estimate the energy injected into the gas to be 1.1-2.0$\times$10$^{56}$~erg, depending on the temperature of the pre-shock gas.

\section{Conclusions}
Using a combination of X-ray and optical emission line data from the \chandra\ and \xmms\ observatories and the GCMS IFU, we have investigated the compact group HCG~57, with a particular focus on the interaction between the two spiral galaxies HCG~57A and D. Prior observations suggest that these galaxies underwent an off-axis collision $\sim$50~Myr ago, with D passing through the disk of A. Based on our observations we draw the following conclusions:

\begin{enumerate}
\item We confirm the presence of an extended hot IGrM extending at least 320~kpc from the group core and with a typical temperature of $\sim$1~keV. Although the temperature declines in the central $\sim$50~kpc, the central cooling time is relatively long (5.9$\pm$0.8~Gyr) and the entropy profile is consistent with the r$^{1.1}$ scaling expected for a halo only weakly affected by cooling and non-gravitational heating. HCG~57 therefore appears to be a system which, while gravitationally bound and relatively relaxed, is only slowly thermally evolving, with cooling from the IGrM unlikely to have any significant impact on the central galaxies in the near future.

\item We find diffuse X-ray emission in 6 of the 8 main member galaxies in the group, likely arising from a mix of unresolved stellar sources (as in HCG~57F), star-formation (important in HCG~57D) and the hot ISM (the dominant component in the largest elliptical, HCG~57C). We also find point-like emission sources in the nuclei of HCG~57A, B, D and possibly E, all of which have previously been identified as optical AGN. The nuclear sources in galaxies A and D are sufficiently bright to allow spectral characterization and we find them to have power-law spectra with photon indices $\Gamma$=1.78$^{+0.37}_{-0.35}$ and 1.65$^{+0.34}_{-0.33}$ respectively. The nucleus of HCG~57A shows evidence of Compton-thin absorption (N$_H$=2.47$^{+0.76}_{-0.71}$$\times$10$^{22}$\pcmsq), which may be occurring in the galaxy disk, as a dust lane is visible adjacent to the nucleus. The nuclei of HCG~57C and E, while too faint for spectral fitting, are found to have fluxes and hardness ratios consistent with a thermal plasma, suggesting that hot gas and/or the stellar population may be a significant contributor to their emission.

\item We identify a further 10 point sources within the \Dtf\ ellipses of the member galaxies. If these are indeed individual sources within the galaxies, all meet the luminosity criterion for classification as ULXs, with L$_{0.5-7 keV}$ in the range 1.3$\times$10$^{39}$-2.3$\times$10$^{40}$\ergps. Statistically we expect 1-2 of these to be chance associations with background sources. Four of the ten sources are associated with the star-forming ring of HCG~57D. A fifth is located close to or within a clump of optical/infared emission northeast of the bulge of HCG~57A, along the line of its interaction with HCG~57D. This source is one of 4 for which forward modelling of the flux and hardness ratio suggests a degree of intrinsic absorption.

\item A bridge of hot X-ray emitting plasma and warm ionized gas is found to link HCG~57D with the core of HCG~57A. The hot component of this bridge is $\sim$10\arcs\ ($\sim$6.4~kpc) wide and 15\arcs\ (9.6~kpc) long, with a temperature of 1.03$^{+0.30}_{-0.41}$~keV. We estimate that the bridge contains $\sim$10$^8$\Msol\ of hot plasma, and that its cooling time (assuming X-ray emission as the dominant radiative process) is long, $\sim$4~Gyr. Emission from ionized gas (H$\alpha$, H$\beta$, [\ion{O}{3}], \nii) is seen throughout the disk of HCG~57D, and in the same bridge region as the X-ray emission linking that galaxy to HCG~57A. Line emission from the bridge is seen at a wide range of velocities (+50 to +350\kmps, relative to the adopted systemic velocity of HCG~57 A+D, 8850\kmps). Line diagnostics show that while much of the disk emission is dominated by star formation, on the west side of the bridge and at the edges of the HCG~57D disk the ionized gas has line ratios consistent with shock heating. The line profiles in the shocked regions also show strong asymmetry, with tails toward shorter wavelengths indicative of gas at a range of velocities.

\item We estimate the velocity of the collision shock responsible for heating the X-ray emitting plasma in the bridge to have been either $\sim$650\kmps\ if the pre-shock medium was cold neutral hydrogen, or $\sim$750\kmps (equivalent to $\mathcal{M}$$\simeq$2.3) if the pre-shock medium was hot ($\sim$0.4~keV). This implies that the collision occurred 30-35~Myr ago, and injected $\sim$10$^{56}$~erg into the hot plasma. Comparison of the optical line diagnostics with shock models suggests that the ionized gas has been subject to slower shocks with velocities $\sim$200-500\kmps. Such variation in shock velocities is expected, given that the pre-shock ISM in the two disks will have been highly multi-phase and that the rotational velocities in the two disks will have increased and decreased the effective interaction velocity in different regions.
\end{enumerate}

The HCG~57A/D collision shows many parallels with similar systems studied in more detail, such as the Taffy galaxies and Stephan's Quintet. This shows the importance of investigating such objects; while not commonplace, collisions between galaxies or between galaxies and tidal gas structures are inevitable in the galaxy group environment, which brings galaxies into close proximity at low relative velocities. These interactions drive the evolution of both the galaxy population and of the wider group, and we should therefore take advantage of the clearest examples to study the complex physics involved. For HCG~57, we now have a relatively clear view of the warm and (thanks to the exceptional spatial resolution of \chandra) hot phases of the gas involved in the interaction, but further observations are required if we are to understand its impact fully. In the Taffy galaxies, radio continuum observations reveal the magnetic field of the intergalactic bridge and the relativistic particles trapped within it. Similar observations of HCG~57 could provide further information on the star formation in the member galaxies while allowing us to examine the bridge structure at higher spatial resolutions than is possible with current data. Prior observations have also shown the interacting galaxies to be rich in molecular gas. Millimeter-wave interferometry offers the possibility of mapping the cold molecular clouds we can expect to find in the bridge via their CO line emission, while \textit{JWST} may offer the ability to track the shock-heated molecular gas via the H$_2$ lines. We hope to see such observations adding to our understanding of galaxy development in the group environment.

\begin{acknowledgments}
We thank the anonymous referee for their positive and helpful comments on the paper. Support for this work was provided by the National Aeronautics and Space Administration through Chandra Award Number GO8-19112A issued by the Chandra X-ray Center, which is operated by the Smithsonian Astrophysical Observatory for and on behalf of the National Aeronautics Space Administration under contract NAS8-03060. PNA would like to thank Guillermo Blanc (Carnegie Observatories) for important assistance in the initial calibration and data reduction for the GCMS. 
\end{acknowledgments}

\facilities{CXO, XMM, Smith (GCMS/Virus-P)} 
\software{CIAO \citep[v4.13][]{Fruscioneetal06}, Sherpa \citep[v4.13][]{Freemanetal01}, SAS \citep[v19.1][]{Gabrieletal04}, Xspec \citep[12.11.0m][]{Arnaud96}, VACCINE \citep{Adams2011}, LZIFU \citep{Ho2016}.}

\bibliographystyle{aasjournal}
\bibliography{paper.bib}

\end{document}